\pdfoutput=1 

\documentclass[12pt]{article}
\usepackage{graphicx}
\usepackage{ifthen} 
\newboolean{pdflatex}
\setboolean{pdflatex}{true} 

\newboolean{articletitles}
\setboolean{articletitles}{true} 

\newboolean{uprightparticles}
\setboolean{uprightparticles}{false} 

\newboolean{inbibliography}
\setboolean{inbibliography}{false} 

\usepackage{afterpage}

\newcommand\pubnumber{DPF2015-457}
\newcommand\pubdate{\today}

\def\napoli{Department of Physics\\
Syracuse University, Syracuse, N.Y. 13244, USA}
\def\support{\footnote{Work supported by U.S. National Science Foundation}}

\def\Title#1{\begin{center} {\Large #1 } \end{center}}
\def\Author#1{\begin{center}{ \sc #1} \end{center}}
\def\Address#1{\begin{center}{ \it #1} \end{center}}

\newcommand\pubblock{\rightline{\begin{tabular}{l} \pubnumber\\
         \pubdate  \end{tabular}}}
\newenvironment{Abstract}{\begin{quotation}  }{\end{quotation}}
\newenvironment{Presented}{\begin{quotation} \begin{center} 
             PRESENTED AT\end{center}\bigskip 
      \begin{center}\begin{large}}{\end{large}\end{center} \end{quotation}}
\def\Acknowledgments{\bigskip  \bigskip \begin{center} \begin{large}
             \bf ACKNOWLEDGMENTS \end{large}\end{center}}

\usepackage{microtype}
\usepackage{lineno}  
\usepackage{xspace} 
\usepackage{caption} 

\usepackage{graphicx}  
\usepackage{color}
\usepackage{colortbl}
\graphicspath{{./figs/}} 

\usepackage{amsmath} 
\usepackage{amssymb}
\usepackage{amsfonts}
\usepackage{upgreek} 

\newcommand*\patchAmsMathEnvironmentForLineno[1]{%
\expandafter\let\csname old#1\expandafter\endcsname\csname #1\endcsname
\expandafter\let\csname oldend#1\expandafter\endcsname\csname
end#1\endcsname
 \renewenvironment{#1}%
   {\linenomath\csname old#1\endcsname}%
   {\csname oldend#1\endcsname\endlinenomath}%
}
\newcommand*\patchBothAmsMathEnvironmentsForLineno[1]{%
  \patchAmsMathEnvironmentForLineno{#1}%
  \patchAmsMathEnvironmentForLineno{#1*}%
}
\AtBeginDocument{%
\patchBothAmsMathEnvironmentsForLineno{equation}%
\patchBothAmsMathEnvironmentsForLineno{align}%
\patchBothAmsMathEnvironmentsForLineno{flalign}%
\patchBothAmsMathEnvironmentsForLineno{alignat}%
\patchBothAmsMathEnvironmentsForLineno{gather}%
\patchBothAmsMathEnvironmentsForLineno{multline}%
\patchBothAmsMathEnvironmentsForLineno{eqnarray}%
}

\usepackage{hyperref}    
\usepackage[all]{hypcap} 

\def\MagUp {\mbox{\em Mag\kern -0.05em Up}\xspace}

\ifthenelse{\boolean{uprightparticles}}%
{

 \def\Pmu         {\ensuremath{\upmu}\xspace}

 \def\Ppsi        {\ensuremath{\uppsi}\xspace}

 \def\PDelta      {\ensuremath{\Delta}\xspace}                 
 \def\PXi      {\ensuremath{\Xi}\xspace}                 
 \def\PLambda      {\ensuremath{\Lambda}\xspace}                 
 \def\PSigma      {\ensuremath{\Sigma}\xspace}                 
 \def\POmega      {\ensuremath{\Omega}\xspace}                 
 \def\PUpsilon      {\ensuremath{\Upsilon}\xspace}                 
 
 \def\PB      {\ensuremath{\mathrm{B}}\xspace}                 
                  
 \def\PD      {\ensuremath{\mathrm{D}}\xspace}

 \def\PJ      {\ensuremath{\mathrm{J}}\xspace}                 
 \def\PK      {\ensuremath{\mathrm{K}}\xspace}

 \def\Pb      {\ensuremath{\mathrm{b}}\xspace}

 \def\Pi      {\ensuremath{\mathrm{i}}\xspace}

 \def\Ps      {\ensuremath{\mathrm{s}}\xspace}

}
{

 \def\Pmu         {\ensuremath{\mu}\xspace}

 \def\Ppsi        {\ensuremath{\psi}\xspace}                 
                  
 \mathchardef\PDelta="7101
 \mathchardef\PXi="7104
 \mathchardef\PLambda="7103
 \mathchardef\PSigma="7106
 \mathchardef\POmega="710A
 \mathchardef\PUpsilon="7107
                  
 \def\PB      {\ensuremath{B}\xspace}                 
                  
 \def\PD      {\ensuremath{D}\xspace}

 \def\PJ      {\ensuremath{J}\xspace}                 
 \def\PK      {\ensuremath{K}\xspace}

 \def\Pb      {\ensuremath{b}\xspace}

 \def\Pi      {\ensuremath{i}\xspace}

 \def\Ps      {\ensuremath{s}\xspace}

}

\makeatletter
\ifcase \@ptsize \relax
  \newcommand{\miniscule}{\@setfontsize\miniscule{4}{5}}
\or
  \newcommand{\miniscule}{\@setfontsize\miniscule{5}{6}}
\or
  \newcommand{\miniscule}{\@setfontsize\miniscule{5}{6}}
\fi
\makeatother

\DeclareRobustCommand{\optbar}[1]{\shortstack{{\miniscule (\rule[.5ex]{1.25em}{.18mm})}
  \\ [-.7ex] $#1$}}

\def\mup        {{\ensuremath{\Pmu^+}}\xspace}
\def\mun        {{\ensuremath{\Pmu^-}}\xspace} 

\def\squark    {{\ensuremath{\Ps}}\xspace}

\def\bquark    {{\ensuremath{\Pb}}\xspace}

  \def\Kbar    {{\kern 0.2em\overline{\kern -0.2em \PK}{}}\xspace}

\def\KorKbar    {\kern 0.18em\optbar{\kern -0.18em K}{}\xspace}

  \def\Dbar    {{\kern 0.2em\overline{\kern -0.2em \PD}{}}\xspace}

\def\DorDbar    {\kern 0.18em\optbar{\kern -0.18em D}{}\xspace}

\def\B       {{\ensuremath{\PB}}\xspace}
\def\Bbar    {{\ensuremath{\kern 0.18em\overline{\kern -0.18em \PB}{}}}\xspace}

\def\BorBbar    {\kern 0.18em\optbar{\kern -0.18em B}{}\xspace}
\def\Bz      {{\ensuremath{\B^0}}\xspace}
\def\Bzb     {{\ensuremath{\Bbar{}^0}}\xspace}

\def\Bsb     {{\ensuremath{\Bbar{}^0_\squark}}\xspace}
\def\Bdb     {{\ensuremath{\Bbar{}^0}}\xspace}

\def\jpsi     {{\ensuremath{{\PJ\mskip -3mu/\mskip -2mu\Ppsi\mskip 2mu}}}\xspace}

  \def\Y#1S{\ensuremath{\PUpsilon{(#1S)}}\xspace}

\def\Xires       {{\ensuremath{\PXi}}\xspace}

\def\Lz          {{\ensuremath{\PLambda}}\xspace}
\def\Lbar        {{\ensuremath{\kern 0.1em\overline{\kern -0.1em\PLambda}}}\xspace}
\def\LorLbar    {\kern 0.18em\optbar{\kern -0.18em \PLambda}{}\xspace}

\def\Lb      {{\ensuremath{\Lz^0_\bquark}}\xspace}
\def\Lbbar   {{\ensuremath{\Lbar{}^0_\bquark}}\xspace}

\def\Xib     {{\ensuremath{\Xires_\bquark}}\xspace}

\def\to                 {\ensuremath{\rightarrow}\xspace}

\def\AT#1     {\ensuremath{A_{\mathrm{T}}^{#1}}\xspace}           

\def\C#1      {\ensuremath{\mathcal{C}_{#1}}\xspace}                       
\def\Cp#1     {\ensuremath{\mathcal{C}_{#1}^{'}}\xspace}                    
\def\Ceff#1   {\ensuremath{\mathcal{C}_{#1}^{\mathrm{(eff)}}}\xspace}        
\def\Cpeff#1  {\ensuremath{\mathcal{C}_{#1}^{'\mathrm{(eff)}}}\xspace}       
\def\Ope#1    {\ensuremath{\mathcal{O}_{#1}}\xspace}                       
\def\Opep#1   {\ensuremath{\mathcal{O}_{#1}^{'}}\xspace}                    

\newcommand{\tev}{\ifthenelse{\boolean{inbibliography}}{\ensuremath{~T\kern -0.05em eV}\xspace}{\ensuremath{\mathrm{\,Te\kern -0.1em V}}}\xspace}
\newcommand{\gev}{\ensuremath{\mathrm{\,Ge\kern -0.1em V}}\xspace}
\newcommand{\mev}{\ensuremath{\mathrm{\,Me\kern -0.1em V}}\xspace}
\newcommand{\kev}{\ensuremath{\mathrm{\,ke\kern -0.1em V}}\xspace}
\newcommand{\ev}{\ensuremath{\mathrm{\,e\kern -0.1em V}}\xspace}
\newcommand{\gevc}{\ensuremath{{\mathrm{\,Ge\kern -0.1em V\!/}c}}\xspace}
\newcommand{\mevc}{\ensuremath{{\mathrm{\,Me\kern -0.1em V\!/}c}}\xspace}
\newcommand{\gevcc}{\ensuremath{{\mathrm{\,Ge\kern -0.1em V\!/}c^2}}\xspace}
\newcommand{\gevgevcccc}{\ensuremath{{\mathrm{\,Ge\kern -0.1em V^2\!/}c^4}}\xspace}
\newcommand{\mevcc}{\ensuremath{{\mathrm{\,Me\kern -0.1em V\!/}c^2}}\xspace}

\def\invfb   {\ensuremath{\mbox{\,fb}^{-1}}\xspace}

\def\gsim{{~\raise.15em\hbox{$>$}\kern-.85em
          \lower.35em\hbox{$\sim$}~}\xspace}
\def\lsim{{~\raise.15em\hbox{$<$}\kern-.85em
          \lower.35em\hbox{$\sim$}~}\xspace}

\def\pt         {\mbox{$p_{\rm T}$}\xspace}

\def\tell1  {TELL1\xspace}
\def\ukl1   {UKL1\xspace}

\newcommand{\eg}{\mbox{\itshape e.g.}\xspace}
\newcommand{\ie}{\mbox{\itshape i.e.}\xspace}

\usepackage{cite} 
\usepackage{mciteplus}

\def\beq{\begin{equation}}
\def\eeq#1{\label{#1}\end{equation}}
\def\eeqn{\end{equation}}

\def\beqa{\begin{eqnarray}}
\def\eeqa#1{\label{#1}\end{eqnarray}}
\def\eeqan{\end{eqnarray}}

\let\bar=\overbar

\def\ie{{\it i.e.}}
\def\eg{{\it e.g.}}

\def\Dslash{\not{\hbox{\kern-4pt $D$}}}
\def\dslash{\not{\hbox{\kern-2pt $\del$}}}

\def\msb{{\bar{\ssstyle M \kern -1pt S}}}

\begin{document}
\begin{titlepage}
\pubblock

\vfill
\Title{Pentaquarks and Tetraquarks at LHCb}
\vfill
\Author{ Sheldon Stone\support} 
\Address{\napoli}
\vfill
\begin{Abstract}
Exotic resonant structures found in \Lb and \Bzb decays into charmonium in the LHCb experiment are discussed.
Examination of the $\jpsi p$ system in $\Lb\to\jpsi K^- p$ decays shows two states each of which must be composed of $uudc\overline{c}$ quarks, and thus are called  charmonium pentaquarks.  
Their masses are $4380\pm 8\pm 29$~MeV and $4449.8\pm 1.7\pm 2.5$~MeV,  and their corresponding widths ($\Gamma$) are $205\pm 18\pm 86$ MeV,  and $39\pm 5\pm 19$ MeV.  The preferred $J^P$ assignments
are of opposite parity,  with one state having spin 3/2 and the other 5/2.   Models of  internal binding of the pentaquark states are discussed. Finally,  another mesonic state is discussed, the $Z(4430)^-$ that decays into $\psi' \pi^-$ and was first observed by the Belle collaboration in $\Bz\to \psi' K^+\pi^-$ decays. Using a sample of approximately 25,000 signal events, LHCb determines the $J^{P}$ to be $1^{+}$.

\end{Abstract}
\vfill
\begin{Presented}
DPF 2015\\
The Meeting of the American Physical Society\\
Division of Particles and Fields\\
Ann Arbor, Michigan, August 4--8, 2015\\
\end{Presented}
\vfill
\end{titlepage}
\def\thefootnote{\fnsymbol{footnote}}
\setcounter{footnote}{0}

\section{Introduction}

In 1964 Gell-Mann \cite{GellMann:1964nj}, and separately Zweig \cite{Zweig:1964}, proposed that hadrons were formed from 
fundamental point like fractionally charged objects now called quarks. For most of the last half-century all well established baryon's could be explained by being composed of three quarks and mesons a quark and an anti-quark. However, in the current decade there have been several observations of candidate mesonic states containing two quarks and two anti-quarks, called tetraquarks \cite{Olsen:2014qna,*Pilloni:2015doa}, and now, as described here, the observation of two pentaquark candidate baryon states \cite{Aaij:2015tga}. Such states were anticipated by Gell-Mann and Zweig. Predictions using theoretical mechanisms in Quantum Chromo-Dynamics (QCD) were made first by Jaffe in 1976 \cite{Jaffe:1976ig} for mesons, and others for baryons in 1978 \cite{Strottman:1979qu,Hogaasen:1978jw}. Several pentaquark observations made about ten years ago were all shown to be fallacious \cite{Hicks:2012zz,*Stone:2000an}. 
Thus, the recent observation of two states decaying into $\jpsi p$, charmonium pentaquarks, found in $\Lb\to\jpsi K^- p$ decays  by the LHCb experiment is surprising.

\begin{figure}[b]
\begin{center}
 \includegraphics[width=0.8\textwidth]{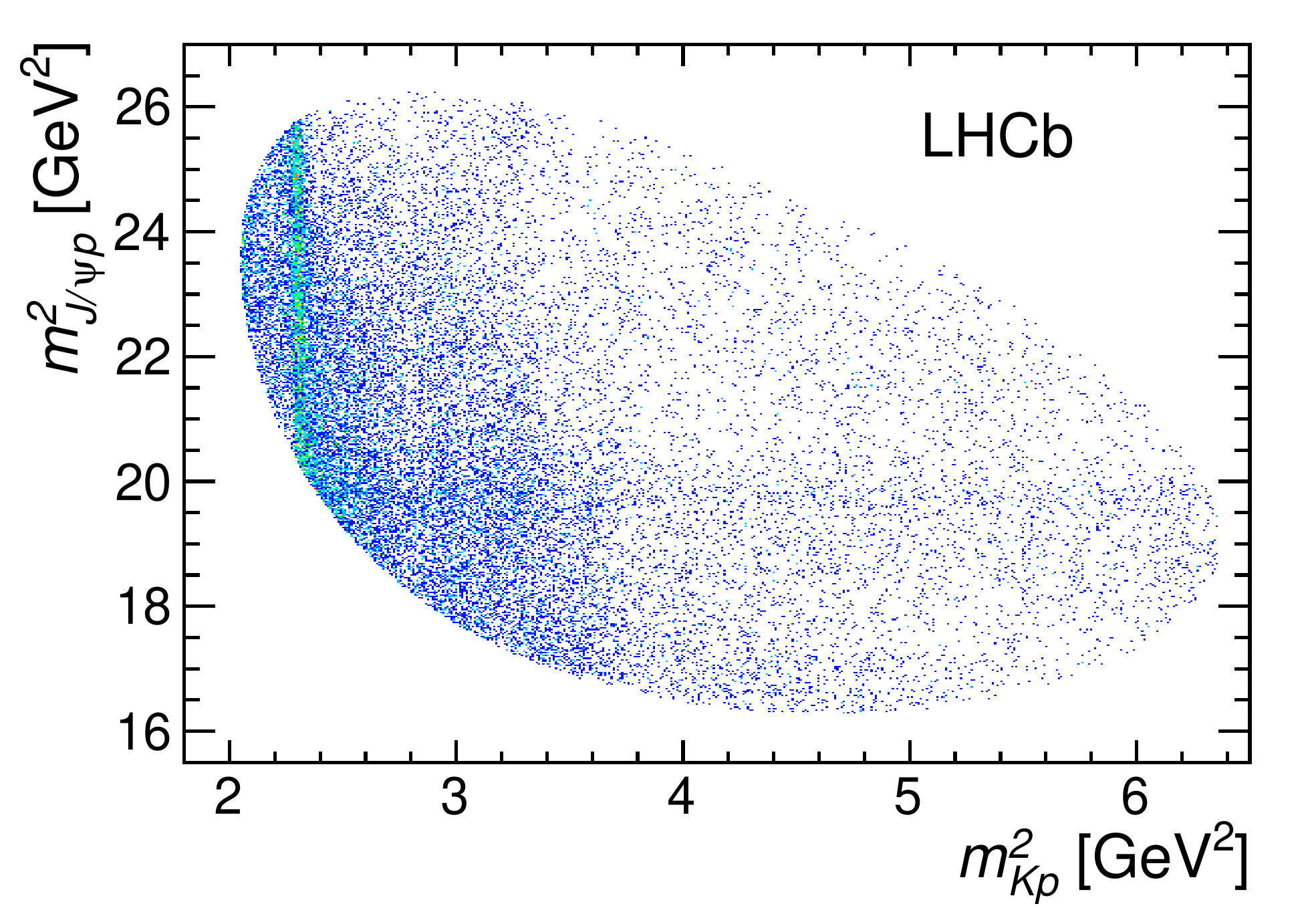}
\end{center}
\vskip -0.7cm
\caption{Invariant mass  squared of $K^-p$  versus $\jpsi p$ for candidates within $\pm15$~MeV of the \Lb mass.}
\label{dlz}
\end{figure}

\begin{figure}[t]
\vskip -0.3cm
\begin{center}
\includegraphics[width=0.46\textwidth]{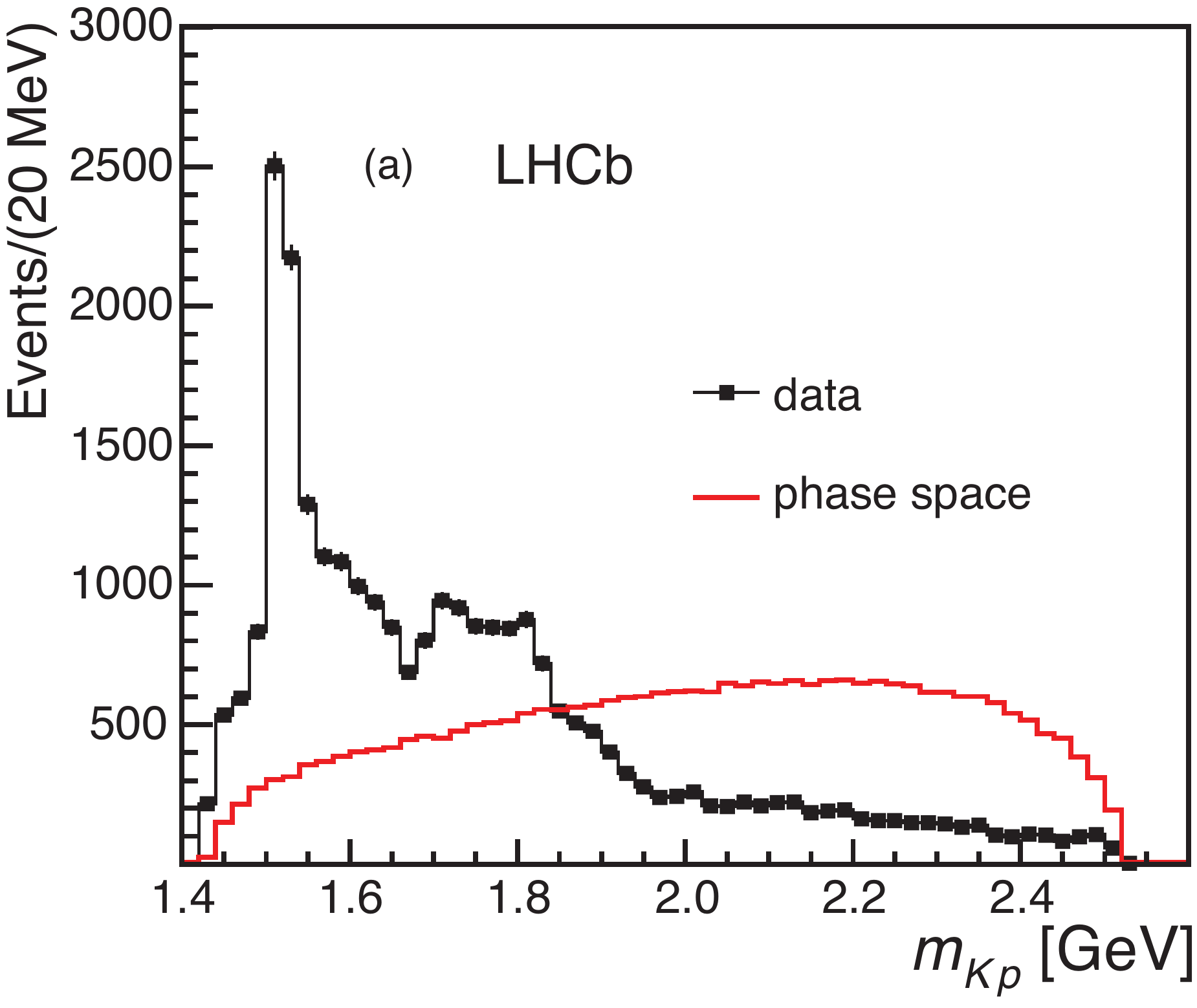}\hspace*{0.5cm}\includegraphics[width=0.46\textwidth]{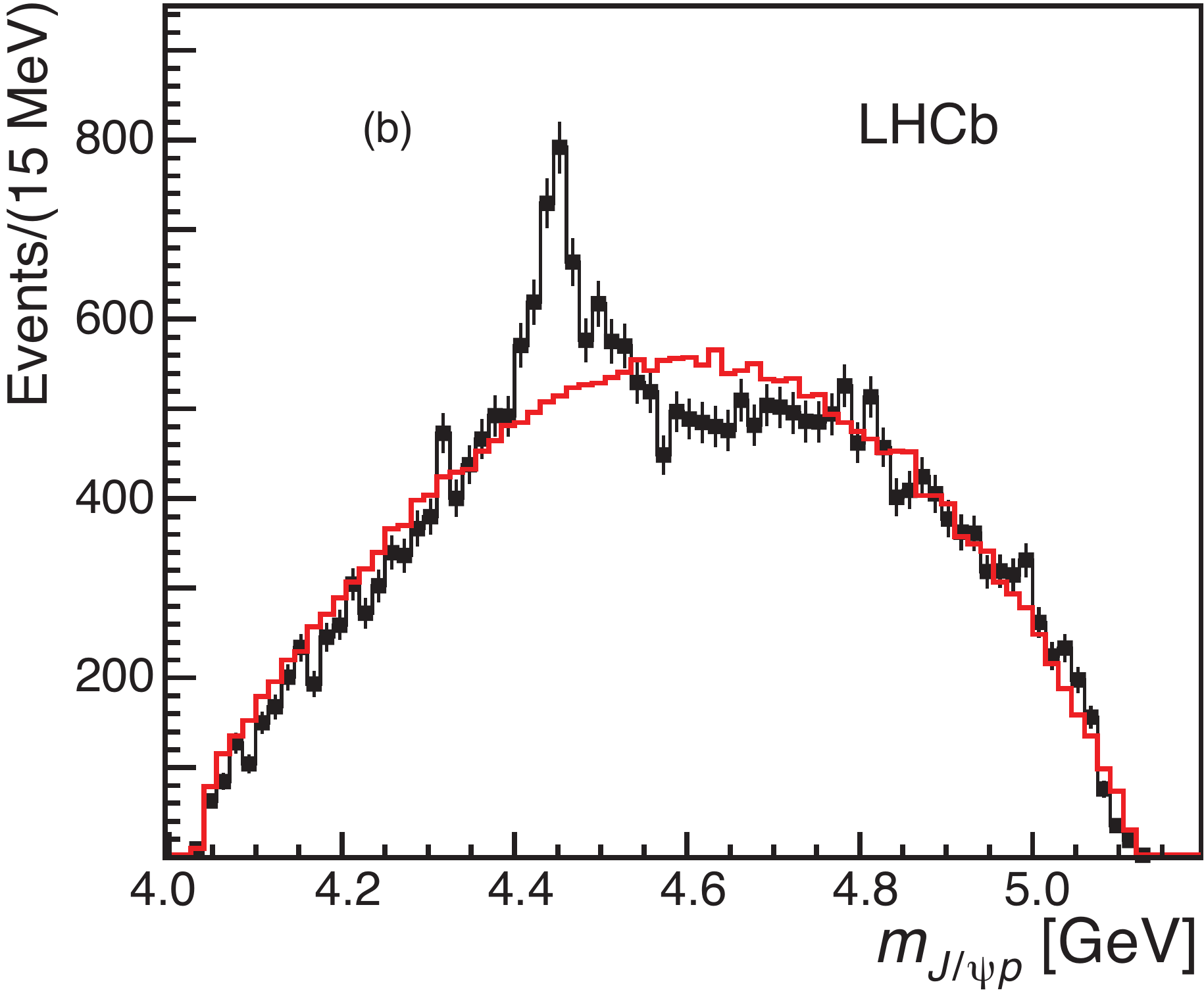}
\end{center}
\vskip -0.4cm
\caption{ Invariant mass of (a) $K^-p$  and (b) $\jpsi p$ combinations from $\Lb\to\jpsi K^-p$ decays. The solid (red) curve is the expectation from phase space. The background has been subtracted.}
\label{mpk-mjpsi}
\end{figure}

The \Lb decay mode was first investigated because it was suggested that it could contribute to the background in suppressed $\Bzb\to\jpsi K^+K^-$ decay that was being searched for and subsequently observed \cite{Aaij:2013mtm}. After its discovery it was used to precisely measure the \Lb baryons lifetime \cite{Aaij:2014zyy,*Aaij:2013oha}. However, one feature of the decay that was not addressed was an anomalous  peaking structure in the $\jpsi p$ invariant mass spectrum, evident in the Dalitz plot shown in Fig.~\ref{dlz}.  While vertical bands correspond to $\Lz^*\to K^-p$ resonances, the horizontal band can only rise from structures in the $\jpsi p$ mass spectrum.
They can also be seen in the invariant mass projections shown in Fig.~\ref{mpk-mjpsi}. 

One may wonder if the peaking in the $\jpsi p$ mass distribution could be caused either by an experimental artifact or by a conspiracy of $\Lz^*$ amplitudes. The investigations described here address these questions.

\section{Analysis and results}

For this study
LHCb \cite{LHCb-det} used data corresponding to 3\invfb of integrated luminosity in 7 and 8~TeV $pp$ collisions. The selection criteria are thoroughly described in the journal article \cite{Aaij:2015tga}. Here I only give a brief summary. Events are kept ({\it{i.e.} triggered upon}) when they contain a  $\jpsi\to\mup\mun$ decay that is detached from the origin of the primary $pp$ collision.

 Track combinations that form $\Lb\to\jpsi K^- p$ candidates are considered if the hadron candidates are positively identified in the RICH system and have significant impact parameters with respect to the primary $pp$ interaction vertex. To reduce backgrounds transverse momentum, \pt, requirements of $>$~500~MeV are imposed on muons and 250~MeV on hadrons.  
Requirements on the \Lb candidate include a vertex $\chi^2<50$  for 5 degrees of freedom,  and a flight distance of greater than 1.5~mm. The vector from the primary vertex to the \Lb vertex must align with the \Lb momentum so that the cosine of the angle between them is larger than $0.999$.  Candidate $\mu^+\mu^-$ combinations are  constrained to the \jpsi mass for subsequent use.

Then a neural network ~\cite{Breiman,*2007physics...3039H} is used to reduce backgrounds while keeping the signal efficiency high. The variables used are the muon identification quality, the probability that both hadron tracks not point at the primary $pp$ collision vertex, the scalar sum of the transverse momentum (\pt) of the two hadrons, and variables related to the \Lb candidate including how well all four tracks form a vertex, the cosine of the angle between a vector from the primary vertex to the  \Lb  vertex and the \Lb momentum vector,  flight distance, and  \pt.

\begin{figure}[b]
\begin{center}
    \includegraphics[width=0.7\textwidth]{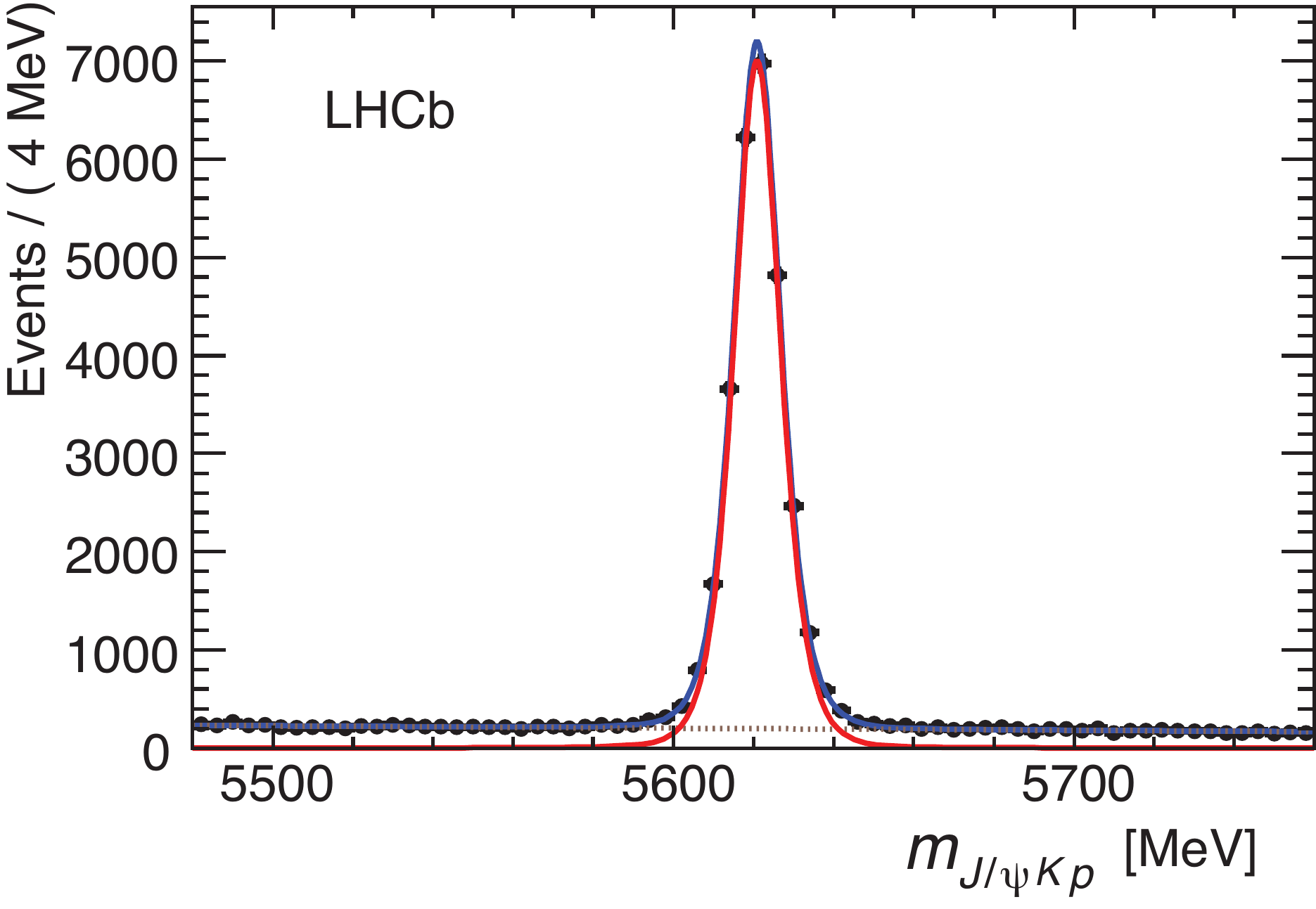} \end{center}
\vskip -0.5cm
\caption{Invariant mass spectrum of $\jpsi K^-p$ combinations, with the total fit, signal and background components shown as solid (blue), solid (red) and dashed lines, respectively.}
\label{fig:rawLb2JpsipK}
\end{figure}

 In addition, specific backgrounds from  $\Bsb$ and $\Bzb$ decays are vetoed. These can occur if the particle identification fails. We remove
  combinations that when interpreted as $\jpsi K^+K^-$ fall within $\pm$30~MeV of the \Bsb mass or when interpreted as $\jpsi K^-\pi^+$ fall within $\pm$30~MeV of the \Bdb mass. This requirement effectively eliminates background from these sources  and causes only smooth changes in the detection efficiencies across the \Lb decay phase space.
Backgrounds from $\Xib$ decays cannot contribute significantly to our sample. The resulting  $\jpsi K^- p$ mass mass spectrum is shown in Fig.~\ref{fig:rawLb2JpsipK}. There are  $26 \,007\pm$166 signal candidates containing 5.4\% background within $\pm 15$~MeV ($\pm2\,\sigma$) of the $\jpsi K^- p$ mass peak.
 For subsequent analysis we constrain the $\jpsi K^- p$ four-vectors to give the \Lb invariant mass and the \Lb momentum vector to be aligned with the measured direction from the primary  to the \Lb vertices \cite{Hulsbergen:2005pu}. 

\begin{figure}[h]
\begin{center}
\includegraphics[width=0.99\textwidth]{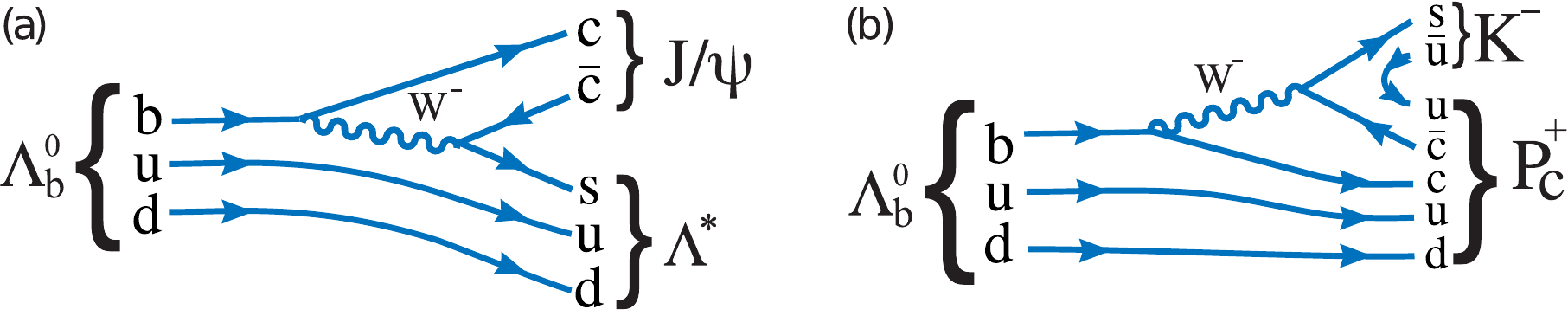}
\end{center}
\vskip -0.3cm
\caption{Feynman diagrams for (a) $\Lb\to \jpsi \Lz^*$ and (b) $\Lb\to P_c^+ K^-$ decay.}
\label{Feynman-Pc}
\end{figure}

\begin{figure}[b]
\begin{center}
\includegraphics[width=.9\textwidth]{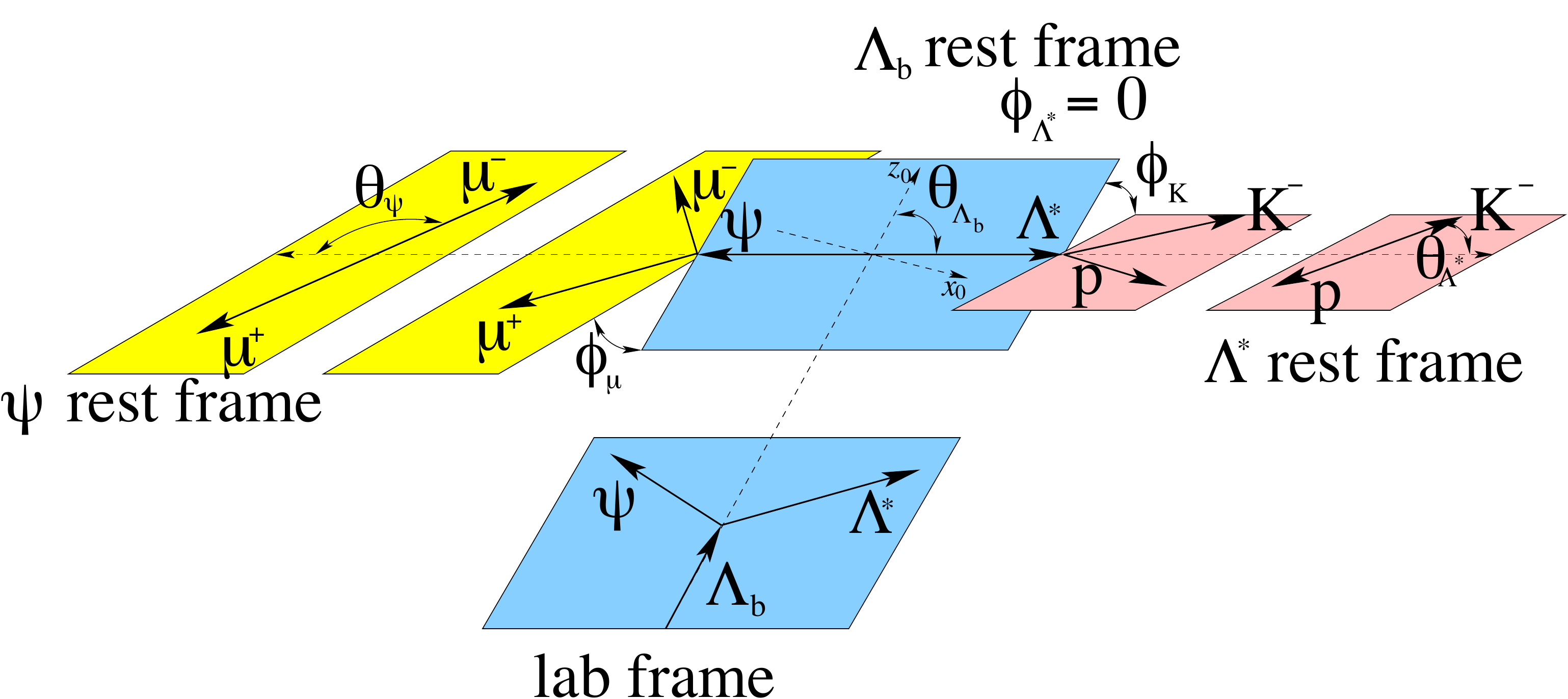}
\end{center}
\vskip -0.1cm
\caption{Definition of the decay angles in the $\Lz^*$ decay chain. }
\label{MatrixElement}
\end{figure}

In this sample specific tracking artifacts were looked for including fake tracks assembled from mismatched upstream and downstream segments, and multiple reconstructions of the same track.
Having found no source of tracking artifacts we proceeded to analyze the decay sequences 
represented by the Feynman diagrams shown in Fig.~\ref{Feynman-Pc}.
This endeavor requires a full analysis of the amplitude for each of the two decay sequences allowing for their mutual interference. The amplitudes are written using six independent variables; one is the invariant $K^-p$ mass, $m_{Kp}$, the others are decay angles. These are shown for the decay sequence $\Lb\to \jpsi \Lz^*;~\Lz^*\to K^-p,~\jpsi\to\mu^+\mu^-$ in Fig.~\ref{MatrixElement}. The $\Lz^*$ resonances are modeled by Breit-Wigner amplitudes except for the $\Lz^*(1405)$ for which a Flatte$^{\prime}$ function is used \cite{Flatte:1976xu}. All other masses, e.g. $m_{\jpsi p}$, and decay angles can be determined from these six quantities.

There are many $\Lz^*$ states that can be considered, and several values of the angular momenta that could be present in each of their decays. Not all of these states are likely to be produced in our final state and not all of the allowable decay angular momenta ($LS$ couplings) are likely to be present. In order to make the most general description possible we first use all the possible states and decay angular momenta; they are listed in Table~\ref{tab:Lstar}.

\begin{table}[htb]
\centering
\caption{The $\Lz^*$ resonances used in the different fits. Parameters are taken from the PDG \cite{PDG}. We take $5/2^-$ for the $J^P$ of the $\Lz(2585)$.
The number of $LS$ couplings is also listed for both the ``reduced'' and ``extended'' models. 
To fix overall phase and magnitude conventions, which otherwise are arbitrary, we fix the lowest angular momentum for the $\Lz(1520)$ decay.  
A zero entry means the state is excluded from the fit. }
\vspace{0.2cm}
\begin{tabular}{lccccc}
\hline\\[-2.5ex] 
State & $J^P$ & $M_0$ (MeV) & $\Gamma_0$ (MeV)& \# Reduced  & \# Extended \\
\hline \\[-2.5ex] 
$\Lz(1405)$ &1/2$^-$ & $1405.1^{+1.3}_{-1.0}$ & $50.5\pm 2.0$ & 3 & 4 \\
$\Lz(1520)$ &3/2$^-$ &$1519.5\pm 1.0$ & $15.6\pm 1.0$& 5 & 6 \\
$\Lz(1600)$ &1/2$^+$ &1600 & 150 &3 & 4 \\
$\Lz(1670)$ &1/2$^-$ & 1670 & 35 & 3 & 4\\
$\Lz(1690)$ &3/2$^-$ & 1690 & 60 & 5 & 6\\
$\Lz(1800)$ &1/2$^-$ & 1800 & 300 &4 &4 \\
$\Lz(1810)$ &1/2$^+$ &1810& 150&3&4\\
$\Lz(1820)$ &5/2$^+$ & 1820 & 80 &1&6\\
$\Lz(1830)$ &5/2$^-$ & 1830 & 95& 1&6 \\
$\Lz(1890)$ &3/2$^+$ & 1890 & 100 &3&6 \\
$\Lz(2100)$ &7/2$^-$ &2100 & 200&1 & 6\\
$\Lz(2110)$ &5/2$^+$ & 2110 & 200  &1 &6\\
$\Lz(2350)$ &9/2$^+$ & 2350 & 150 &0 &6\\
$\Lz(2585)$ &?& $\approx$2585 & 200 &0 & 6\\\hline
\end{tabular}
\label{tab:Lstar}
\end{table}

\begin{figure}[b]
\begin{center}
\vskip 8mm
\includegraphics[width=0.50\textwidth]{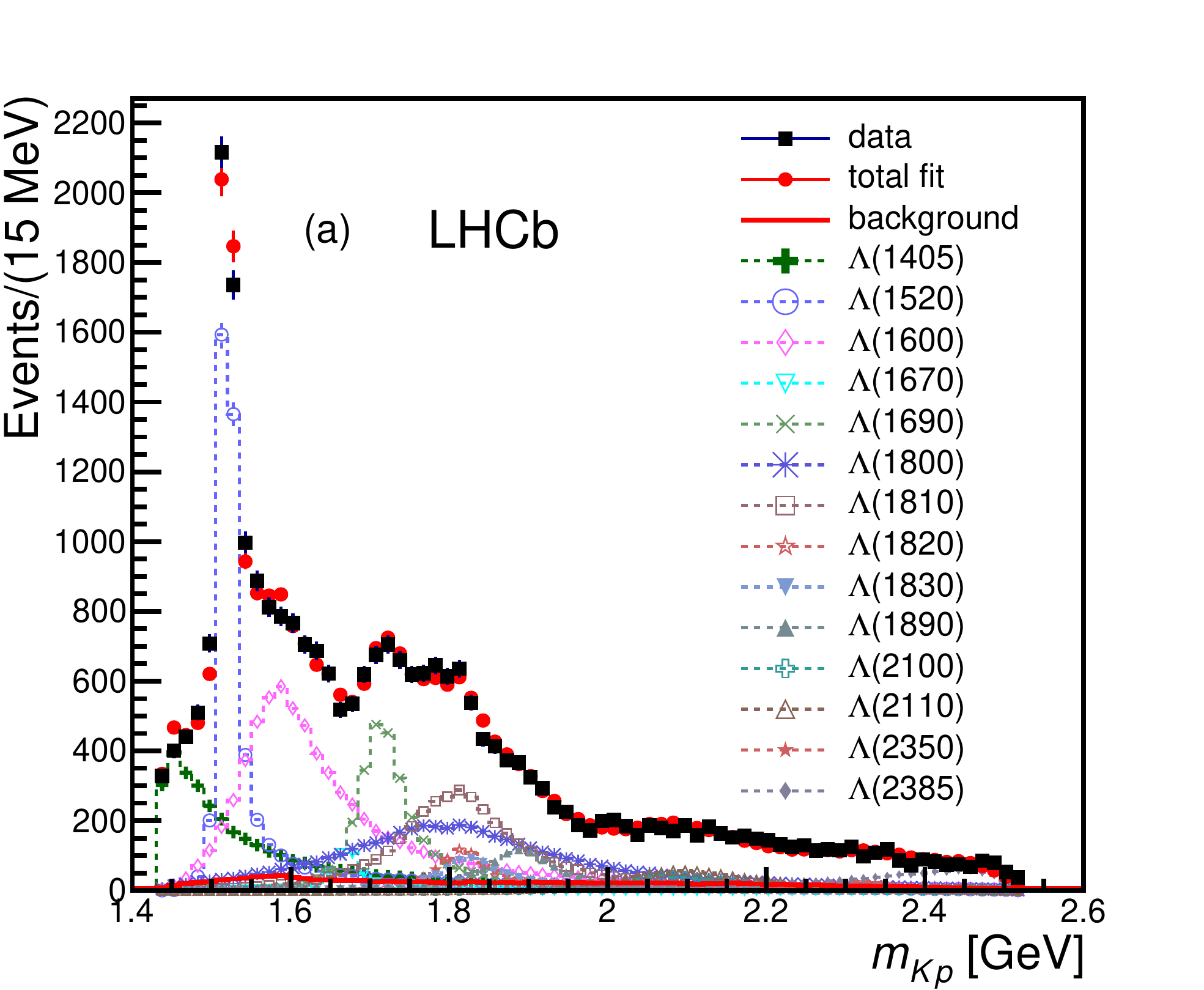}\includegraphics[width=0.50\textwidth]{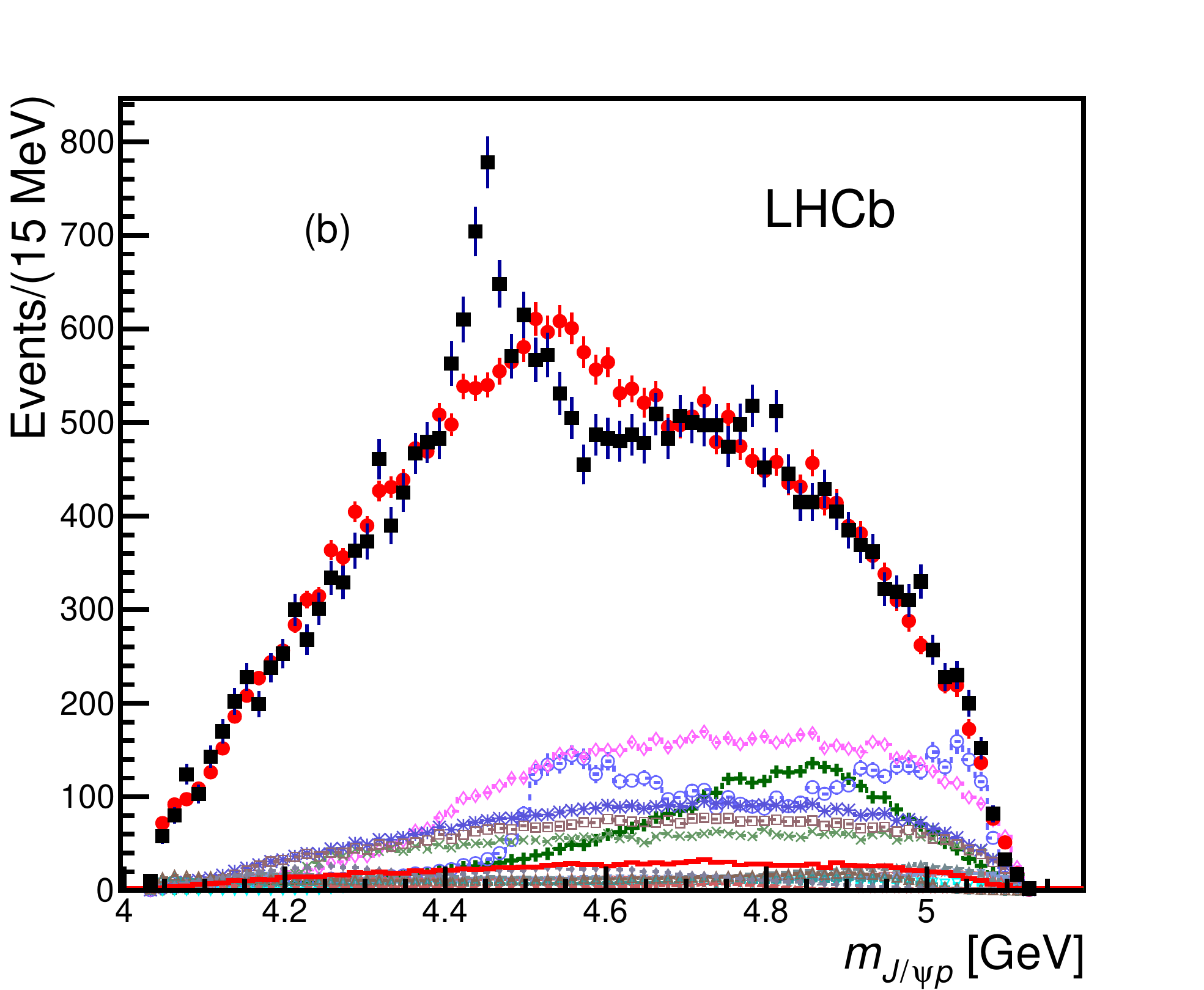}
\end{center}
\vskip -0.5cm
\caption{Results for (a) $m_{Kp}$  and (b) $m_{\jpsi p}$ for the extended $\Lz^*$  model fit without $P_c^+$ states. The data are shown as (black) squares with error bars, while the (red) circles show the results of the fit. The error bars on the points showing the fit results are due to simulation statistics.}
\label{Mall}
\end{figure}

Then data are then fit to this model which has 146 free helicity couplings, even with the masses and widths of the resonant states fixed to their PDG values, done in order to allow the fit to converge. (Variations are considered later as part of the systematic uncertainties.) The results of the fit are shown in Fig.~~\ref{Mall}. The fit gives a good description of the $\Lz^*$ states as can be seen in the $m_{Kp}$ spectrum but fails miserably to reproduce the structure in $m_{\jpsi p}$. 

Not satisfied with using all the known $\Lz^*$ states we tried several other different configurations: (i) we added all the possible $\PSigma^*$ states, (ii) we added two additional $\Lz^*$ allowing their masses and widths to float in the fit and allowed spins up to $5/2$ with both parities, and (iii) we added four non-resonant components with $J^P=1/2^+,~1/2^-,3/2^+,$ and $3/2^-$. None of these fits explains the data, indeed the improvements were small.

\begin{figure}[t]
\begin{center}
\hbox{\includegraphics[width=.95\textwidth]{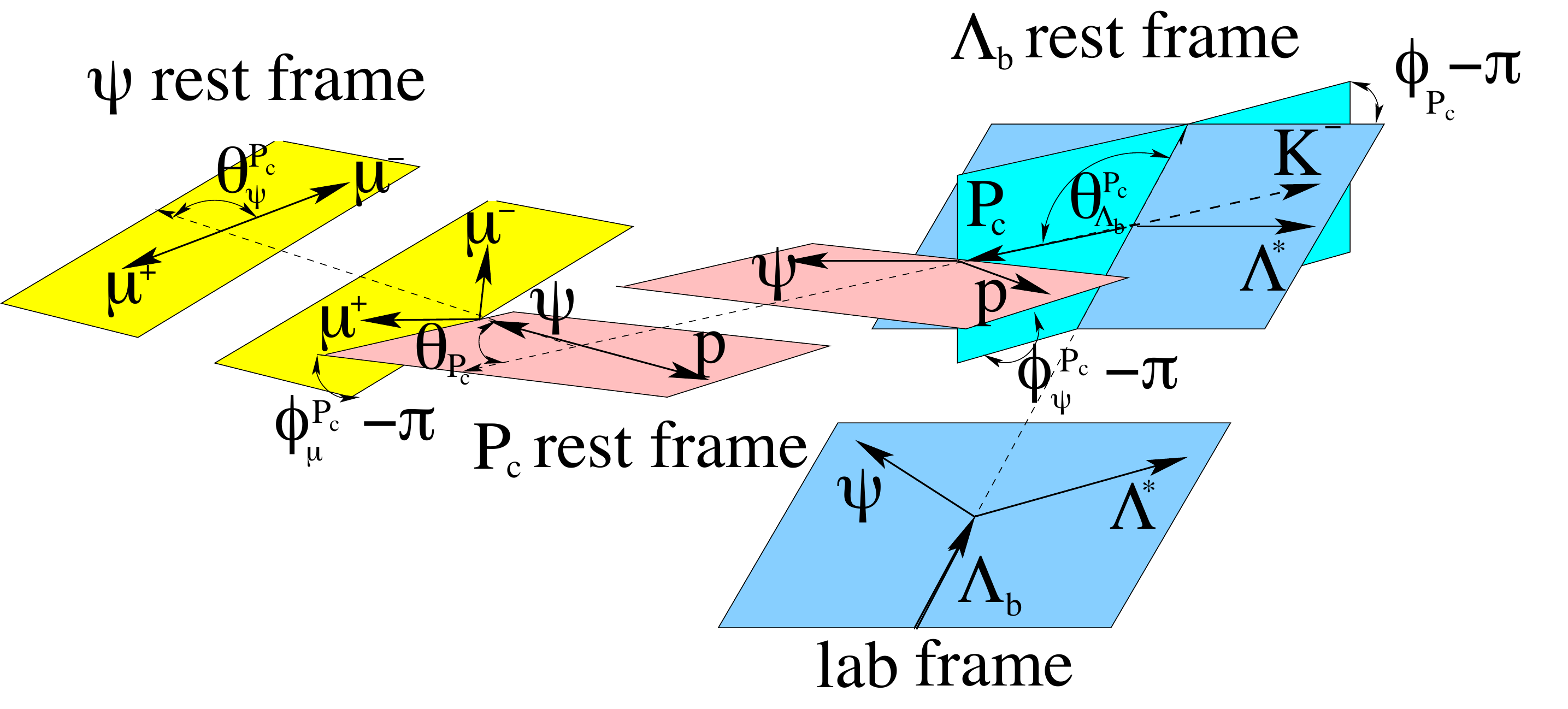}\hskip-4cm\quad}
\end{center}
\vskip -0.4cm
\caption{
Definition of the decay angles in the $P_c^+$ decay sequence. 
}
\label{fig:helicitypc}
\end{figure}

Having failed to describe the data without a resonant state decaying into $\jpsi p$ we added one. Now to do this we must write the matrix element for the decay sequence $\Lb\to P_c^+ K^-,~P_c^+\to \jpsi p$ in terms of the same decay angular variables as the previous decay sequence involving only $\Lz^*$ decays. The decay angles before the appropriate rotations that put the decays in the same rest frames are shown in Fig.~\ref{fig:helicitypc}. The derivation of the matrix element in full mathematical detail is given in the arXiv article and the supplementary material for the Physical Review Letters publication \cite{Aaij:2015tga}.

In each fit we minimize  $-2\ln{\cal{L}}$ where ${\cal{L}}$ represents the fit likelihood.
The difference of $\Delta\equiv-2\ln{\cal{L}}$ between different
amplitude models reflects the goodness of fit.
For two models representing separate hypotheses, \eg\, when discriminating between
different $J^P$ values assigned to a $P_c^+$ state, 
the assumption of a $\chi^2$ distribution with one degree of freedom
for $\Delta$ under the disfavored $J^P$ hypothesis 
allows the calculation of a lower limit on the significance of its 
rejection, \ie\, the p-value \cite{james2006statistical}.
Therefore, it is convenient to express  values of $\Delta$ as
${n^2_\sigma}$, where $n_\sigma$ corresponds to the  number of  standard
deviations in the normal distribution with the same p-value. 
When discriminating between models without and with $P_c^+$ states,
$n_\sigma$ overestimates the p-value by a modest amount. Thus, we use
simulations to obtain better estimates of the significance of the
$P_c^+$ states.

\begin{figure}[htb]
\begin{center}
\includegraphics[width=0.49\textwidth]{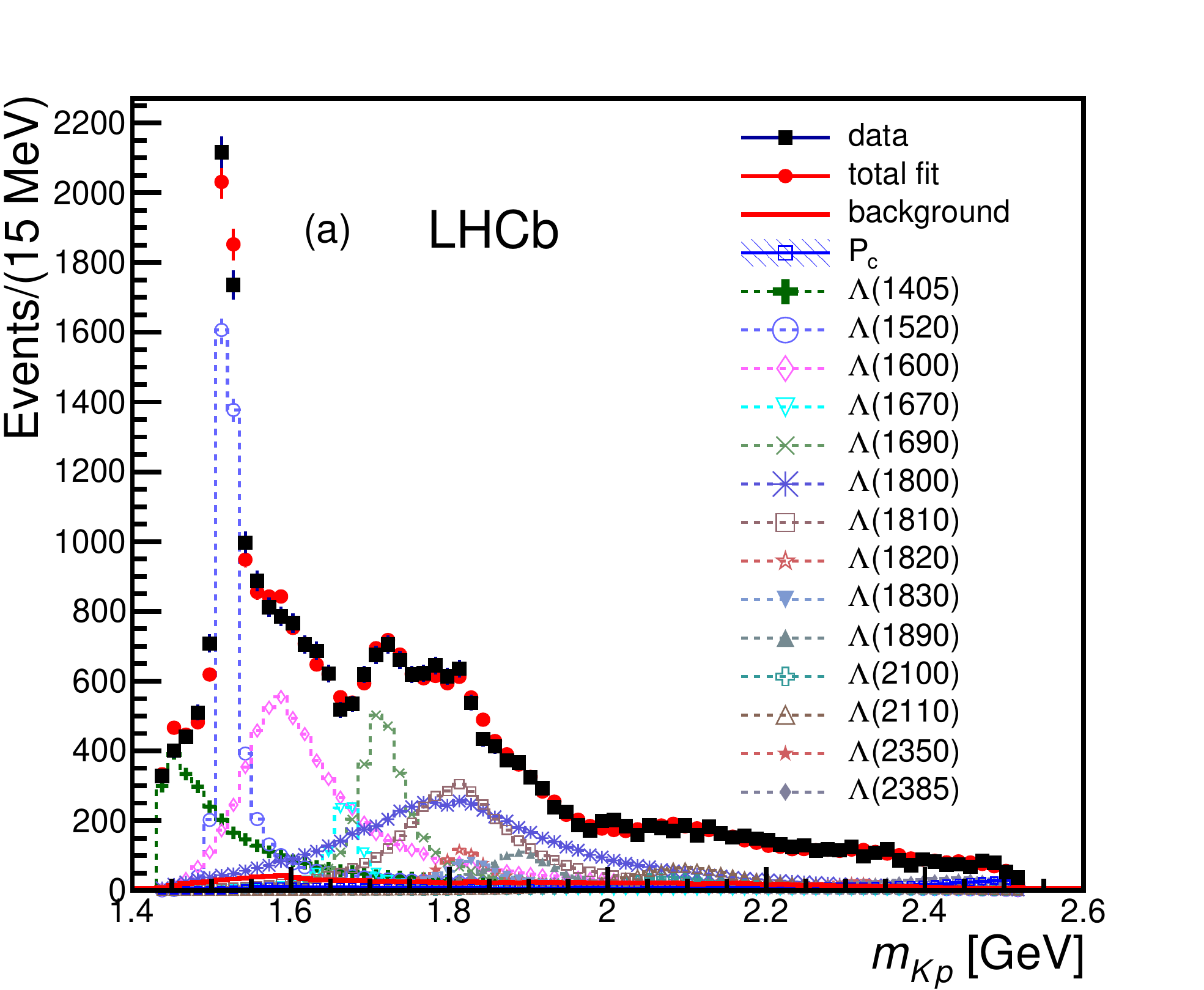}
\includegraphics[width=0.49\textwidth]{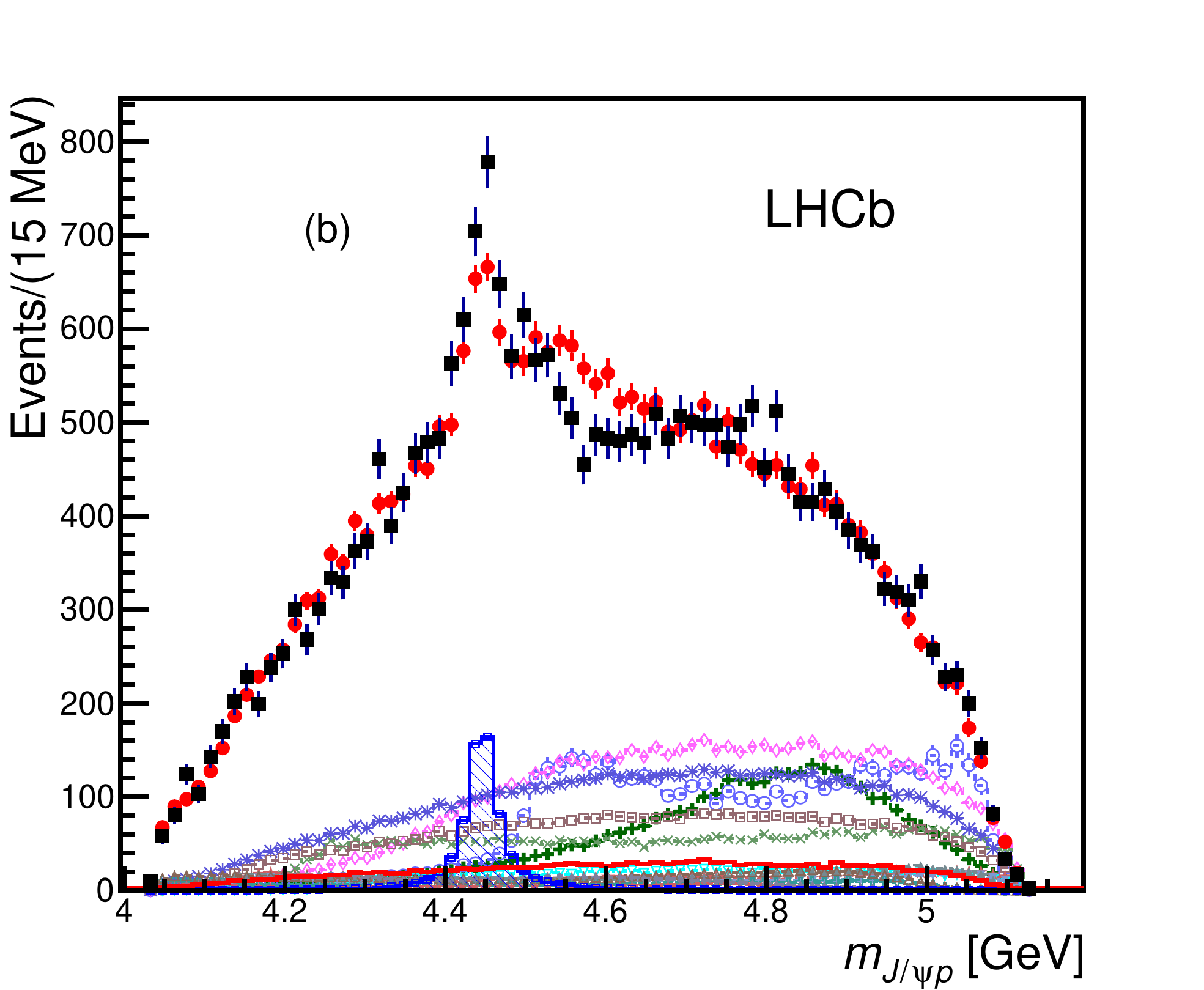}
\end{center}
\vskip -0.1cm
\caption{Results of the fit with one $J^P=5/2^+$ $P_c^+$ candidate. (a) Projection of the invariant mass of $K^-p$ combinations from $\Lb\to\jpsi K^-p$ candidates. The data are shown as (black) squares with error bars, while the  (red) circles show the results of the fit;  (b) the corresponding $\jpsi p$ mass projection. The (blue) shaded plot shows the $P_c^+$ projection, the other curves represent individual  $\Lz^*$ states. }
\label{Extended-1Pc}
\end{figure}

We perform separate fits for 
$J^P$ values of $1/2^{\pm}$, $3/2^{\pm}$ and $5/2^{\pm}$. The mass and width of the putative $P_c^+$ state are allowed to vary. The best fit prefers a $5/2^+$ state, which improves $-2\ln {\cal{L}}$ by 215. 
Figure~\ref{Extended-1Pc} shows the projections for this fit.  While the $m_{Kp}$ projection is well described, clear discrepancies in $m_{\jpsi p}$ remain visible. 

\begin{figure}[b!]
\begin{center}
\includegraphics[width=0.49\textwidth]{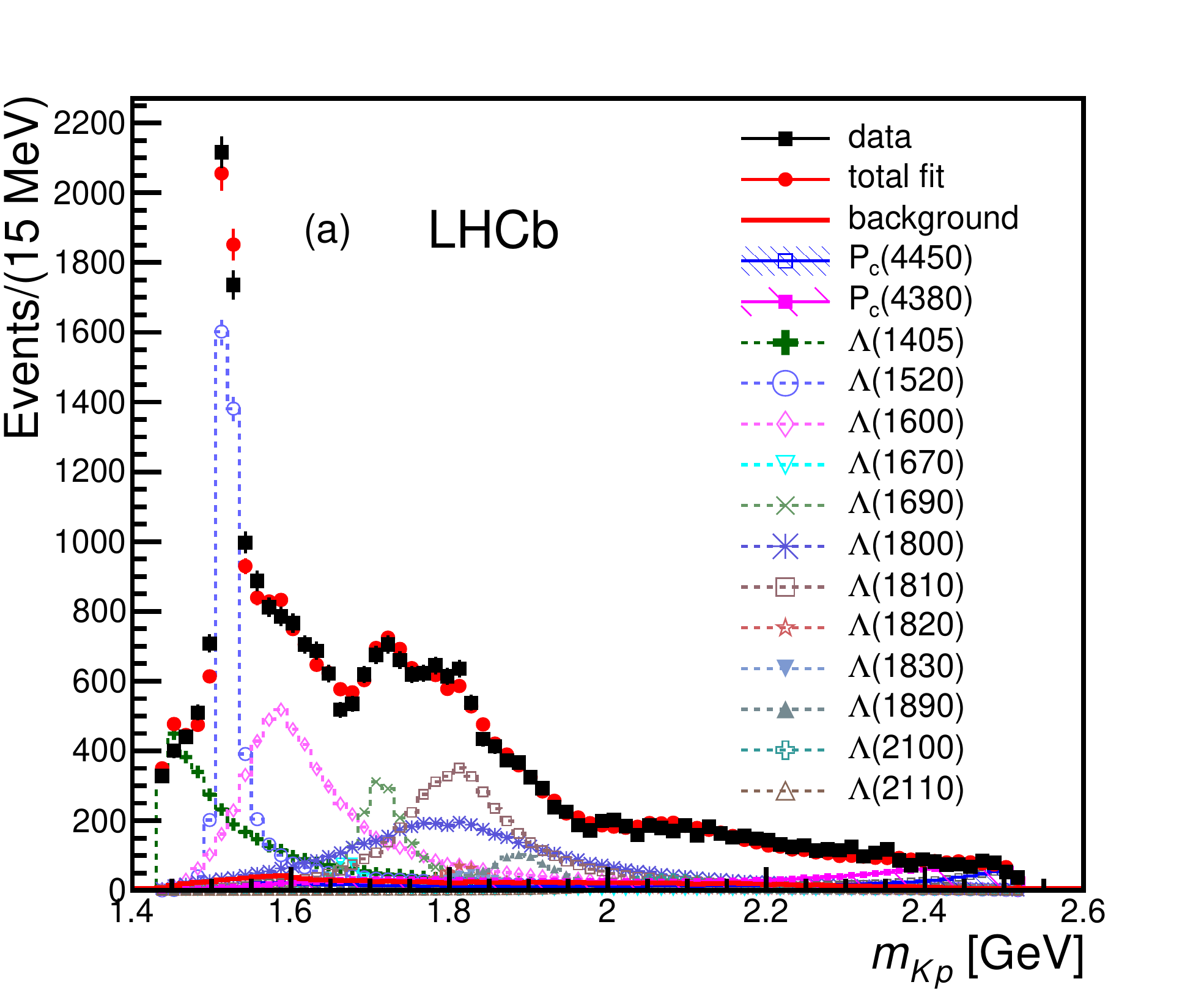}\includegraphics[width=0.49\textwidth]{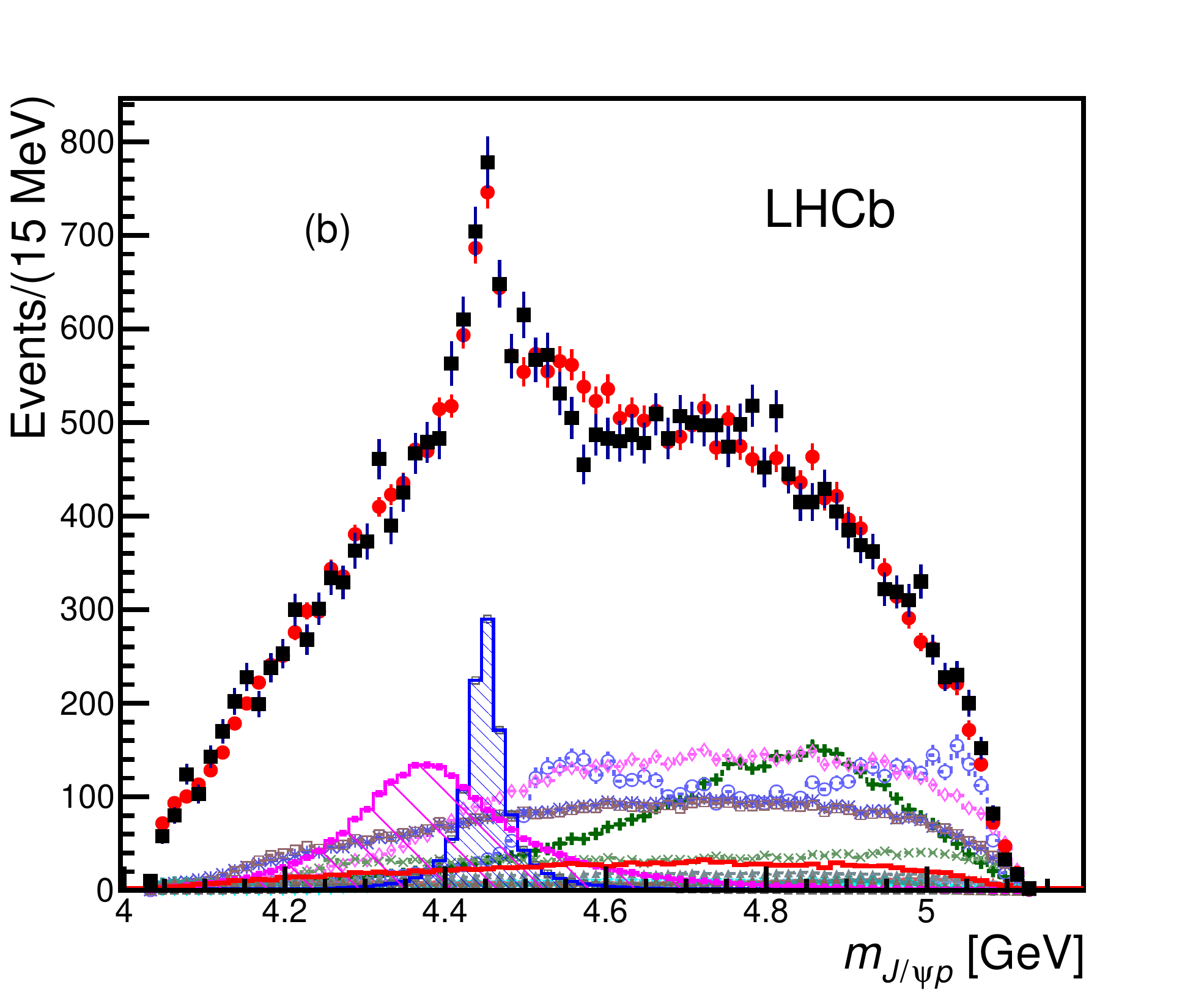}
\end{center}
\vskip -0.1cm
\caption{Fit projections for (a) $m_{Kp}$  and (b) $m_{\jpsi p}$ for the reduced $\Lz^*$ model with two $P_c^+$ states (see Table~\ref{tab:Lstar}). The data are shown as solid (black) squares, while the solid (red) points show the results of the fit.  The solid (red) histogram shows the background distribution. The (blue) open squares with the shaded histogram represent the $P_c(4450)^+$ state, and the shaded histogram topped with (purple) filled squares represents the $P_c(4380)^+$ state. Each $\Lz^*$ component is also shown.}
\label{Pc2d}
\end{figure}

The next step is to fit with two $P_c^+$ states including their allowed interference. These fits were performed both with the reduced model and the extended model in order to estimate systematic uncertainties.  Toy simulations are done to more accurately evaluate the statistical significances of the two states, resulting in  9 and 12 standard deviations, for  lower mass and higher mass states, using the extended model which gives lower significances. The best  fit projections are shown in Fig.~\ref{Pc2d}. Both  $m_{Kp}$ and the peaking structure in $m_{\jpsi p}$ are reproduced by the fit. The reduced model has 64 free parameters for the $\Lz^*$ rather than 146 and allows for a much more efficient examination of the parameter space and, thus, is used for numerical results.
The two  $P_c^+$ states are found to have masses of  $4380\pm 8\pm 29$~MeV and $4449.8\pm 1.7\pm 2.5$~MeV, with corresponding widths of  $205\pm 18\pm 86$ MeV and $39\pm 5\pm19$ MeV. (Whenever two uncertainties are quoted the first is statistical and the second systematic.)  
The fractions of the total sample due to the lower mass and higher mass states are ($8.4\pm0.7\pm4.2$)\% and ($4.1\pm0.5\pm 1.1)$\%, respectively. The overall branching fraction has recently be determined to be  \cite{Aaij:2015fea}
\begin{equation}
{\cal{B}}(\Lb\to \jpsi K^- p)= \left(3.04\pm 0.04^{+0.55}_{-0.43}\right)\times 10^{-4},
\end{equation}
where the systematic uncertainty is largely due to the normalization procedure, leading to the product branching fractions:
\begin{align}
{\cal{B}}(\Lb\to P_c(4380)^+ K^- p){\cal{B}}(P_c(4380)^+ \to \jpsi p)= &\left(2.56^{+1.38}_{-1.34}\right)\times 10^{-5} \nonumber \\
{\cal{B}}(\Lb\to P_c(4450)^+ K^- p){\cal{B}}(P_c(4450)^+ \to \jpsi p)= &\left(1.25^{+0.42}_{-0.40}\right)\times 10^{-5},
\end{align} 
where all the uncertainties have been added in quadrature.

\begin{figure}[htb]
\begin{center}
\includegraphics[width=1.0\textwidth]{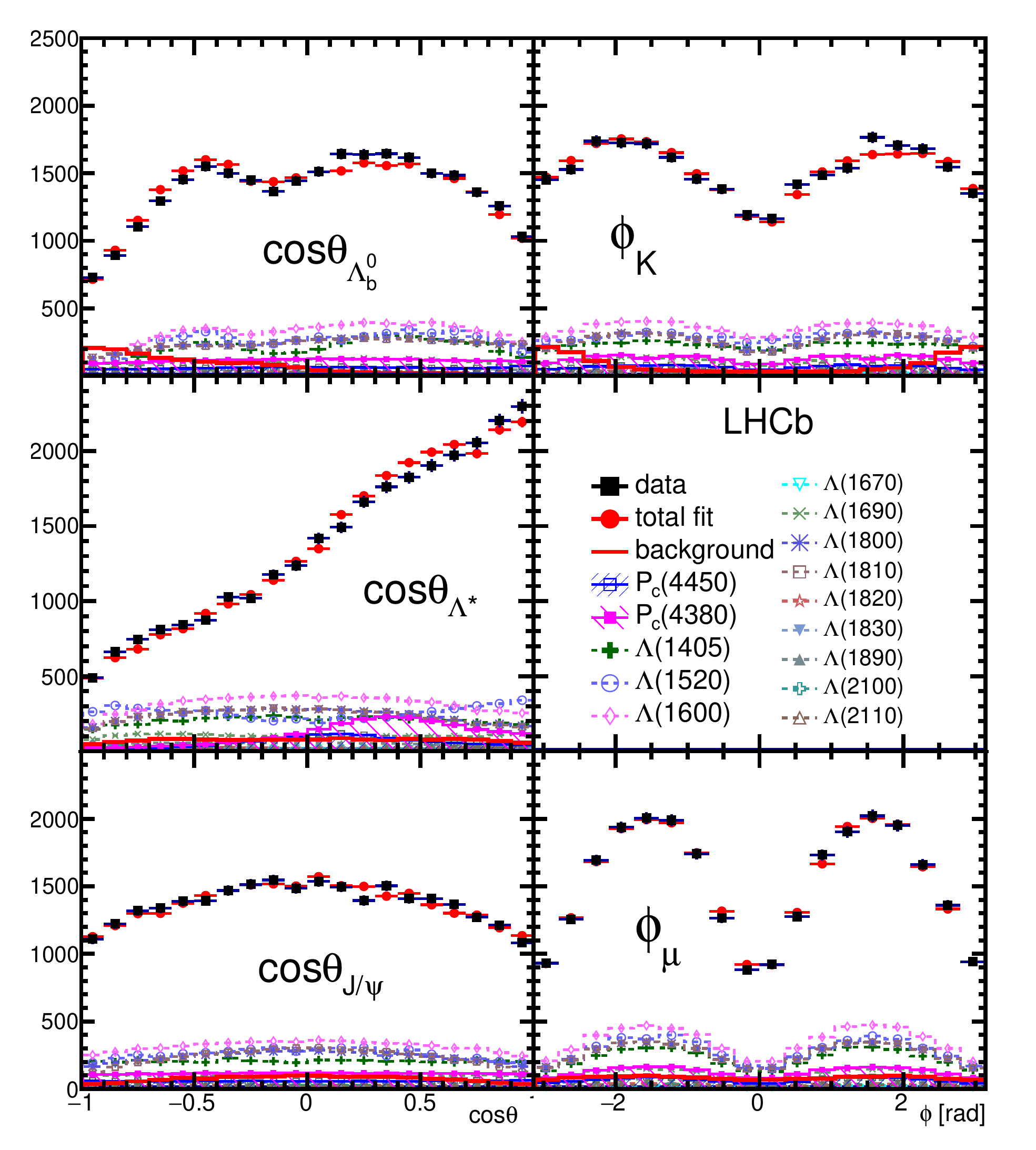}
\end{center}
\vskip -0.1cm
\caption{Various decay angular distributions for the fit with two $P_c^+$ states. The data are shown as (black) squares, while the (red) circles show the results of the fit.  Each fit  component is also shown. The angles are defined in the text.}
\label{TwoZ-angular-cFit}
\end{figure}

 The best fit solution has spin-parity $J^P$ values of ($3/2^-$, $5/2^+$). Acceptable solutions are  also found for additional cases with opposite parity, either ($3/2^+$, $5/2^-$) or ($5/2^+$, $3/2^-$). 
 The five angular distributions are also well fit as can be seen in Fig.~\ref{TwoZ-angular-cFit}.

\begin{figure}[H!]
\begin{center}
\includegraphics[width=1.0\textwidth]{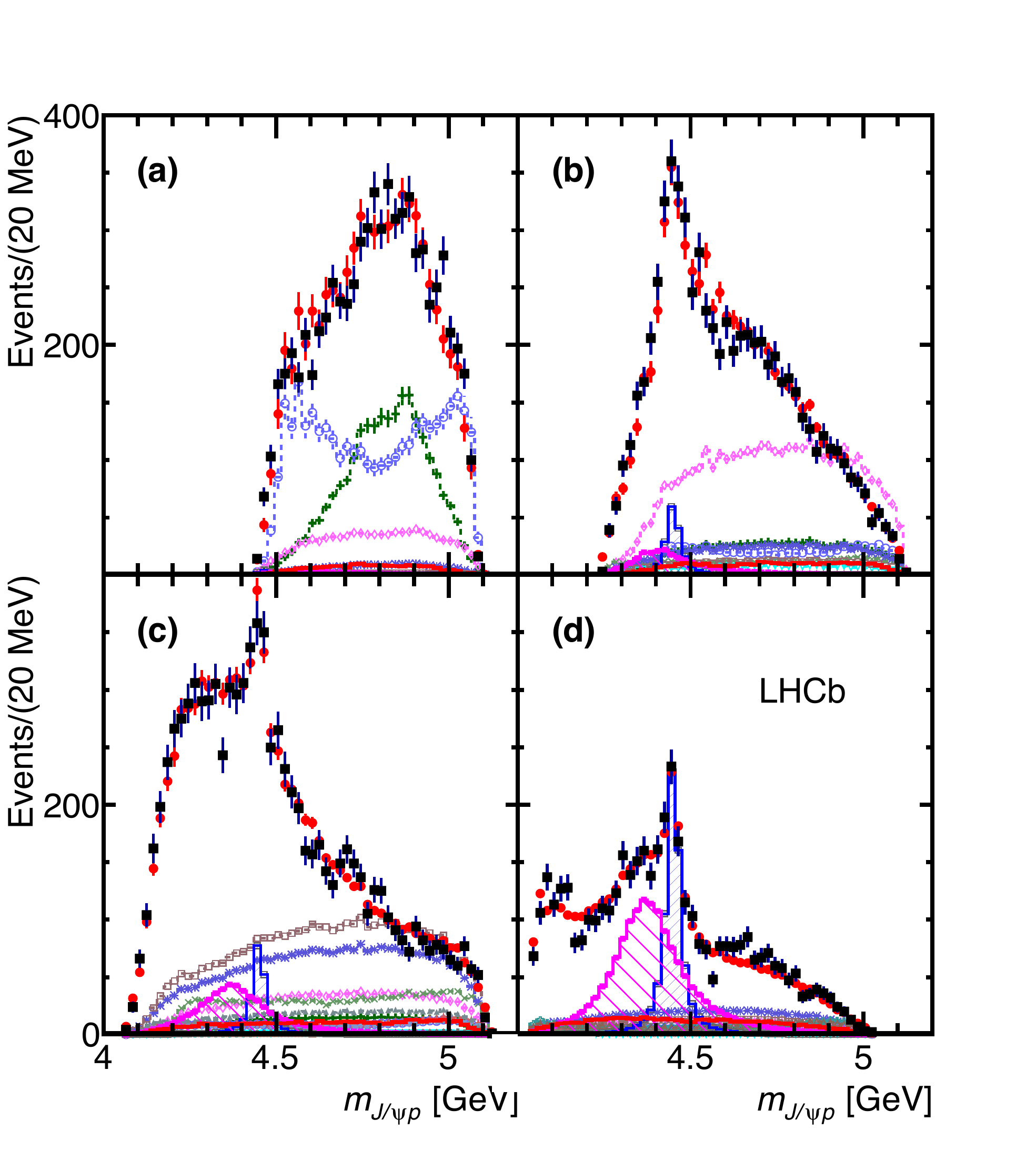}\end{center}
\vskip -0.5cm
\caption{$m_{\jpsi p}$ in various intervals of $m_{Kp}$ for the fit with two $P_c^+$ states: (a) $m_{Kp}<1.55$~GeV, (b) $1.55<m_{Kp}<1.70$~GeV, (c) $1.70<m_{Kp}<2.00$~GeV, and (d) $m_{Kp}>2.00$~GeV.  The data are shown as  (black) squares with error bars, while the (red) circles show the results of the fit. The blue and purple histograms show the two $P_c^+$ states. See Fig.~\ref{TwoZ-angular-cFit} for the legend.}
\label{TwoZ-mjpsip-bins-cFit}
\end{figure}

\begin{figure}[hbt]
\begin{center}
\includegraphics[width=0.55\textwidth]{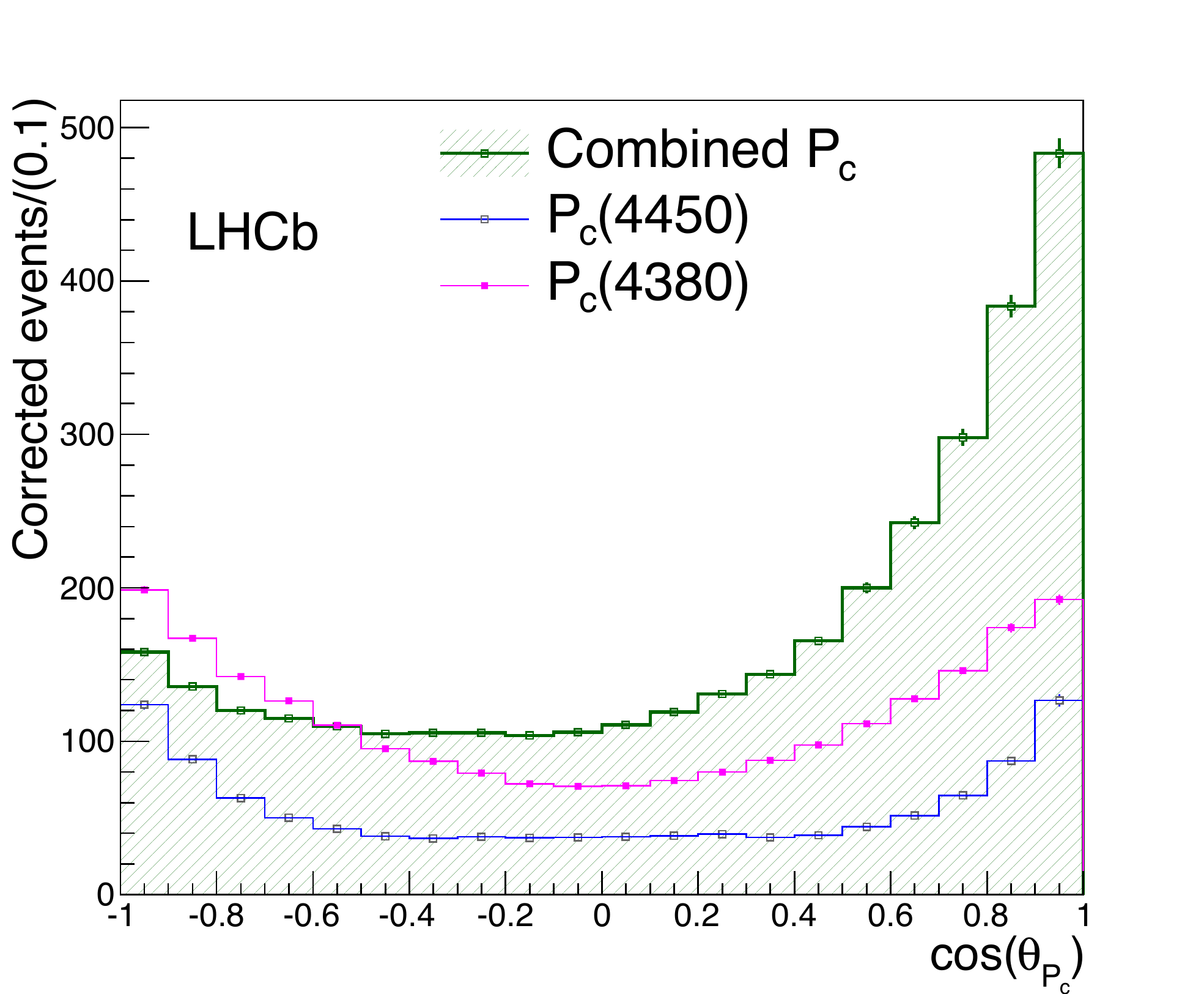}\end{center}
\vskip -0.5cm
\caption{Efficiency corrected and background subtracted fit projections of the decay angular distributions for the two $P_c^+$ states and their sum. Values of $\cos\theta_{P_c}$ near $-1$ are correlated with values of $m_{Kp}$ near threshold, while those near $+1$ are correlated with higher values. }
\label{cosPc_v4}
\end{figure}

The fit projections in different slices of $K^-p$ invariant mass are given in Fig.~\ref{TwoZ-mjpsip-bins-cFit}.  In slice (a) the $P_c^+$ states are not present, nor should they be as they are outside of the Dalitz plot boundary. In slice (d) both $P_c^+$ states form a large part of the mass spectrum; there is also a considerable amount of negative interference between them. This can be seen better by examining the decay angle of the $P_c^+$, $\theta_P$, the angle of the proton in $\jpsi p$ rest frame with respect to the $P_c^+$ direction transformed into its rest frame, shown in Fig.~\ref{cosPc_v4} for the entire $m_{Kp}$ range. The summed fit projections agrees very well with the angular distributions in the data showing that two interfering states are needed to reproduce the asymmetric distribution.\footnote{It can be shown mathematically that the states need to be of opposite parity.}

{\clearpage}

Systematic uncertainties are evaluated for the masses, widths and fit fractions of the $P_c^+$ states, and for the fit fractions of the two lightest and most significant $\Lz^*$ states. Additional sources of modeling uncertainty that we have not considered may affect the fit fractions of the heavier $\Lz^*$ states. The sources of systematic uncertainties  are listed in Table~\ref{tab:syssum}.
They include differences between the results of the extended versus reduced model, varying the $\Lz^*$ masses and widths, uncertainties in the identification requirements for the proton, and restricting its momentum, inclusion of a nonresonant amplitude in the fit, use of separate higher and lower \Lb mass sidebands, alternate $J^P$ fits, varying the Blatt-Weisskopf barrier factor, $d$,  between 1.5 and 4.5~GeV$^{-1}$ in the Breit-Wigner mass shape-function, changing the  angular momentum $L$  by one or two units, and accounting for potential mis-modeling of the efficiencies. For the $\Lz(1405)$ fit fraction we also added an uncertainty for the Flatt\'e couplings, determined by both halving and doubling their ratio, and taking the maximum deviation as the uncertainty.

The stability of the results is cross-checked by comparing the data recorded in 2011/2012, with the LHCb dipole magnet polarity in up/down configurations, \Lb/\Lbbar decays, and \Lb produced with low/high values of \pt. The fitters were tested on simulated pseudoexperiments and no biases were found. In addition, selection requirements are varied, and the vetoes of \Bsb and \Bdb  are removed and explicit models of those backgrounds added to the fit; all give consistent results.

\begin{table}[hbt]
\centering
\caption{Summary of systematic uncertainties on $P_c^+$ masses, widths and fit fractions, and $\Lz^*$ fit fractions. 
A fit fraction is the ratio of the phase space integrals of the matrix element squared for a single resonance and for the total amplitude. 
The terms ``low" and ``high" correspond to the lower and higher mass $P_c^+$ states.}
\vspace{0.2cm}
\resizebox{\textwidth}{!}{ 
\begin{tabular}{lrrrrrrrcc}
\hline
~~~Source & \multicolumn{2}{c}{$M_0$ (MeV)$\!\!\!\!$} &\multicolumn{2}{c}{$\Gamma_0$ (MeV)$\!\!\!\!$} &\multicolumn{4}{c}{Fit fractions (\%)}$\!\!\!\!$\\
 &low& high$\!\!$ & low& high$\!\!$& low& high&$\Lz(1405)$&$\Lz(1520)$\\
\hline
Extended vs. reduced &$21$ &0.2&$54$&10 &$3.14$&0.32&1.37~~~&0.15\\
$\Lz^*$ masses \& widths &7 &0.7 &20&4&0.58&0.37&2.49~~~&2.45\\
Proton ID &2& 0.3 & $1$ & $2$&0.27&0.14&0.20~~~&0.05\\
$10<p_p<100$~GeV &  0&1.2&$1$&$1$&0.09&0.03&0.31~~~&0.01\\
Nonresonant & 3&0.3&$34$&~2 &$2.35$&$0.13$&3.28~~~&0.39\\
Separate sidebands & 0 &~~0&~5&~0&$0.24$&$0.14$&0.02~~~&0.03\\
$J^P$ ($3/2^+$, $5/2^-$) or ($5/2^+$, $3/2^-$) &$10$&$1.2$&34&10&$0.76$&0.44&& \\
$d=1.5-4.5~$GeV$^{-1}$&$9$&0.6&19&$3$&0.29&0.42&0.36~~~&1.91\\
$L_{\Lb}^{{P_c^+}}$ $\Lb\to P_c^+~{\rm (low/high)} K^-$ &6&0.7&4&8&$0.37$&0.16&&\\
$L_{{P_c^+}}$ $P_c^+~{\rm (low/high)}\to\jpsi p$&4&$0.4$&31&7&$0.63$&0.37&&\\
$L_{\Lb}^{\Lz^*_{\!n}}$ $\Lb\to \jpsi \Lz^*$&11&0.3&20&2&0.81&0.53&3.34~~~&2.31\\
Efficiencies &1&0.4&4&0&0.13&0.02&0.26~~~&0.23\\
Change $\Lz(1405)$ coupling&0 &0&0&0&0&0&1.90~~~&~~~0\\
\hline
Overall & 29&2.5&86&19&4.21&1.05&5.82~~~&3.89\\\hline
\end{tabular}
}
\label{tab:syssum}
\end{table}

\begin{figure}[htb]
\begin{center}
\includegraphics[width=0.9\textwidth]{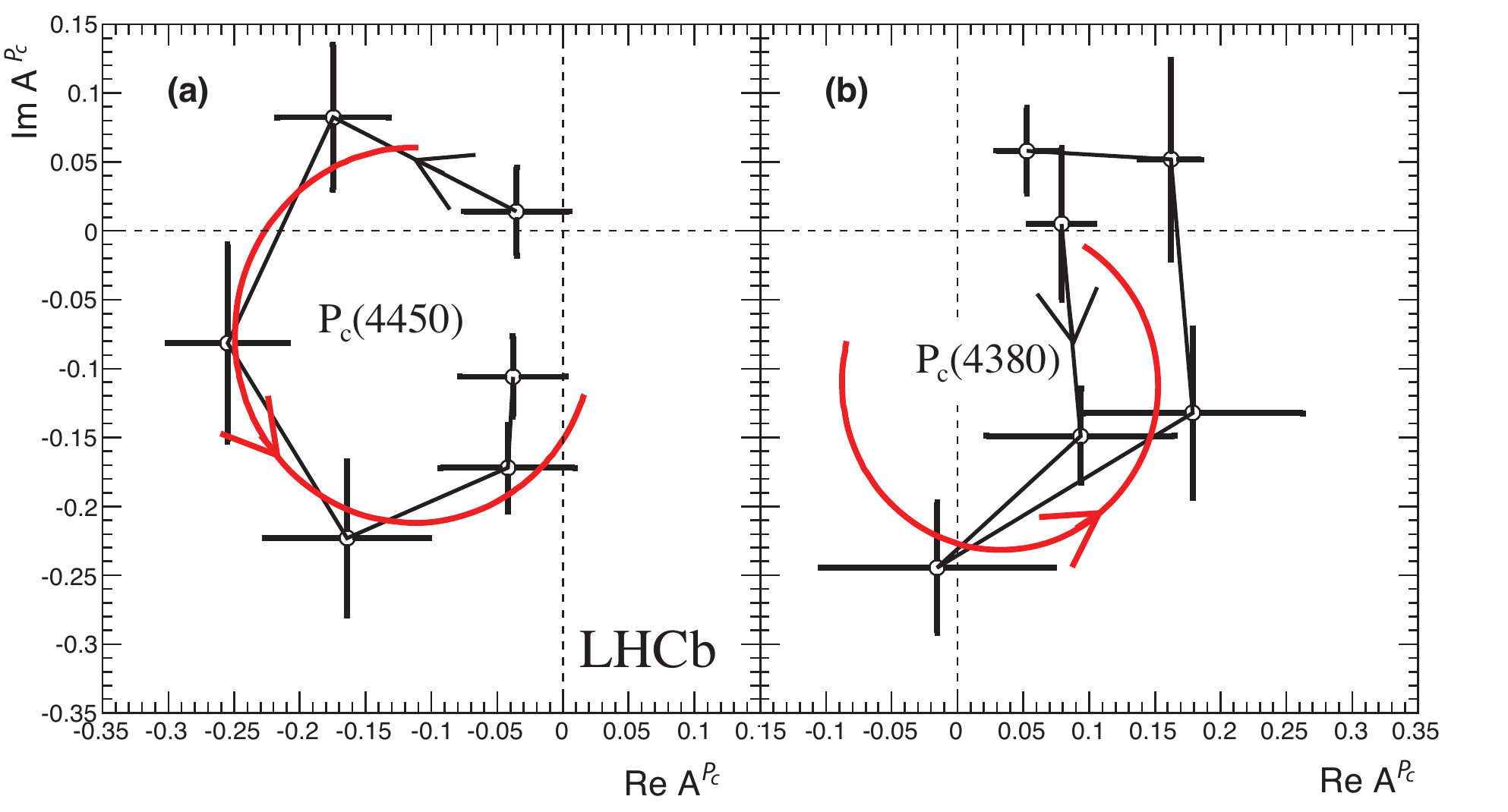}
\vskip -0.2cm
\caption{
Fitted values of the real and imaginary parts of the amplitudes for the baseline ($3/2^-$, $5/2^+$) fit for a) the $P_c(4450)^+$ state and b) the $P_c(4380)^+$ state, each divided into six $m_{\jpsi p}$ bins of equal width between $-\Gamma$ and $+\Gamma$ shown in the Argand diagrams as connected points with error bars ($m_{\jpsi p}$ increases counterclockwise).
The solid (red) curves are the predictions
from the Breit-Wigner formula for the same mass ranges
with  $M$ ($\Gamma$) of
 4450 (39) \mev and 4380 (205) \mev, respectively,
with the phases and magnitudes at the resonance masses set to the
average values between the two points around $M$.  
The phase convention is fixed by the $\Lz(1520)$. Systematic uncertainties are not included.}
\label{DoubleArgand}
\vskip -2 cm
\end{center}
\end{figure}

Further evidence for the resonant character of the higher mass, narrower, $P_c^+$ state is obtained by viewing the evolution of the complex amplitude in the Argand diagram  \cite{PDG}.
In the amplitude fits discussed above, the $P_c(4450)^+$ is represented by a Breit-Wigner amplitude, where
the magnitude and phase vary with $m_{\jpsi p}$ according to
an approximately circular trajectory in the (Re$\,A^{P_c}$, Im$\,A^{P_c}$) plane, where
$A^{P_c}$ is the $m_{\jpsi p}$ dependent part of the $P_c(4450)^+$ amplitude.
We perform an additional fit to the data using the reduced $\Lz^*$ model, in which
we represent the $P_c(4450)^+$ amplitude as the combination of independent
complex amplitudes at six equidistant points in the range $\pm \Gamma=39\,$MeV around $M=4449.8\,$MeV as determined in the default fit.
Real and imaginary parts of the amplitude are interpolated in mass between the fitted points.
The resulting Argand diagram, shown in Fig.~\ref{DoubleArgand}(a),
is consistent with a rapid counter-clockwise change of the $P_c(4450)^+$ phase when its magnitude
reaches the maximum, a behavior characteristic of a resonance. A similar study for the wider state is shown in Fig.~\ref{DoubleArgand}(b); although the fit does show a large phase change, the amplitude values are sensitive to the details of the $\Lz^*$ model  and so this latter study is not conclusive. 

\section{Models of pentaquark structure}
\begin{figure}[b!]
\begin{center}
\includegraphics[width=0.3\textwidth]{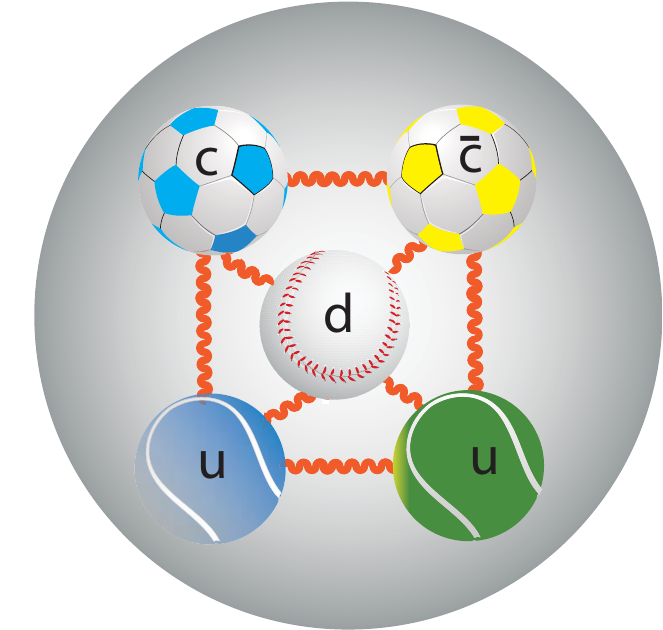}\hspace*{1.5cm}
\includegraphics[width=0.3\textwidth]{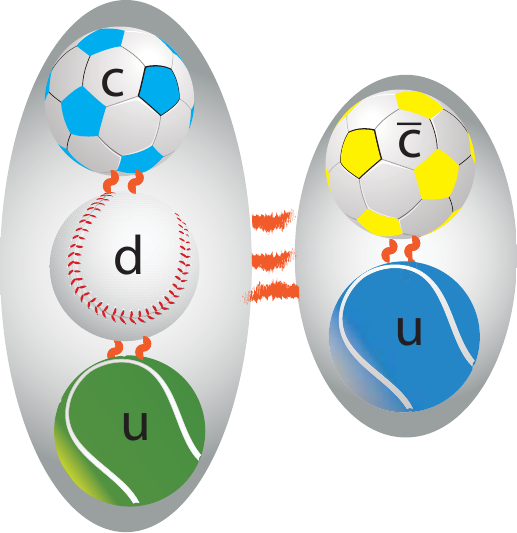}
\vskip -0.02cm
\caption{ (left) Illustration of a tightly bound $P_c^+$ state, and (right) a molecularly bound state.}
\label{bound}
\end{center}
\end{figure}

All models must explain the $J^P$ of the two states not just one. They also should predict properties of other yet to be observed states: masses,
widths, $J^P$'s. There are many explanations of the $P_c^+$ states.  Let us start with tightly bound quarks ala Jaffe \cite{Jaffe:1976ig}. Early work  \cite{Strottman:1979qu,Hogaasen:1978jw,Rossi:1977cy} has been expanded upon recently using diquark models \cite{ Maiani:2015vwa,*Lebed:2015tna,*Anisovich:2015cia,*Li:2015gta,*Ghosh:2015ksa,*Wang:2015epa,*Wang:2015ava}. Here each pair of two quarks form a colored objects along with the lone antiquark. The three colors then form a colorless state, as illustrated in Fig.~\ref{bound}(left).

Molecularly bound states, which also build on previous work \cite{Voloshin:1976ap, *DeRujula:1976qd, *Tornqvist:1991ks, *Tornqvist:1993ng,*Yang:2011wz,*Wang:2011rga}, have recently received much attention. Models trying to explain the states discussed here have already appeared \cite{Karliner:2015ina,*Roca:2015dva,*Chen:2015loa,*He:2015cea,*Meissner:2015mza}, and even been disputed \cite{Mironov:2015ica}. A molecular state configuration is illustrated in Fig.~\ref{bound}(right).

Other attempts at explaining the data are based on concepts of rescattering \cite{Guo:2015umn,*Mikhasenko:2015vca}. These types of models were proposed to explain other resonances such as the $a_1(1260)$. These postdictions are made by constructing  an amplitude that is consistent with the data in shape. They make no prediction of the magnitude of the amplitude, or its  width, nor do they predict other final states where the phenomena could be encountered. Sometimes the phase motion is calculated.  

\begin{figure}[b]
\begin{center}
\includegraphics[width=0.9\textwidth, angle=0.8]{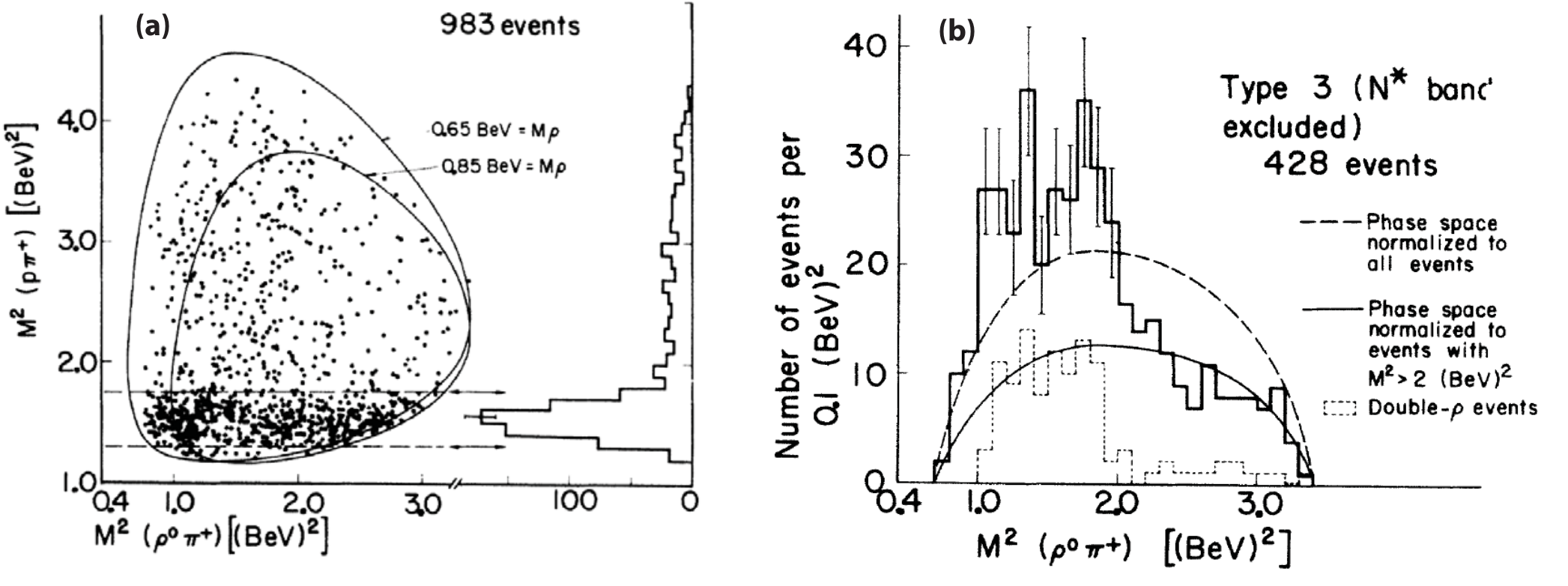}
\vskip -0.02cm
\caption{ (a) Dalitz plot for $\pi^+p\to \pi^+\rho^0$ events. The $N^*$ resonance band is indicated by horizontal dashed lines. (b) Histogram of the invariant $\rho^0\pi^+$ mass-squared for events outside the $N^*$. The peaks correspond in mass to the $a_1^+$ and $a_2^+$ resonances.}
\label{Goldhaber}
\end{center}
\end{figure}

The $a_1(1260)^+$ saga is a good example even if it's 51 years old, indeed as old as the quark model. Track measurements from a bubble chamber experiment using 3.65~GeV incident beam $\pi^+$ mesons that reacted as $\pi^+ p\to \pi^-\pi^+\pi^+$ were analyzed \cite{Goldhaber:1964zz}. After restricting the data to have one $\pi^+\pi^-$ mass combination consistent with the $\rho^0$ mass they obtained  the Dalitz plot shown in Fig.~\ref{Goldhaber}(a). Removing events in the low mass $p \pi^-$ band, due to $N^*$ resonances, they were left with the resulting $\rho^0\pi^+$ mass spectrum on the right. This can be explained by a higher mass $a_2^+$ state and a new $a_1^+$ state at lower mass. 

\begin{figure}[htb]
\begin{center}
\includegraphics[width=0.9\textwidth]{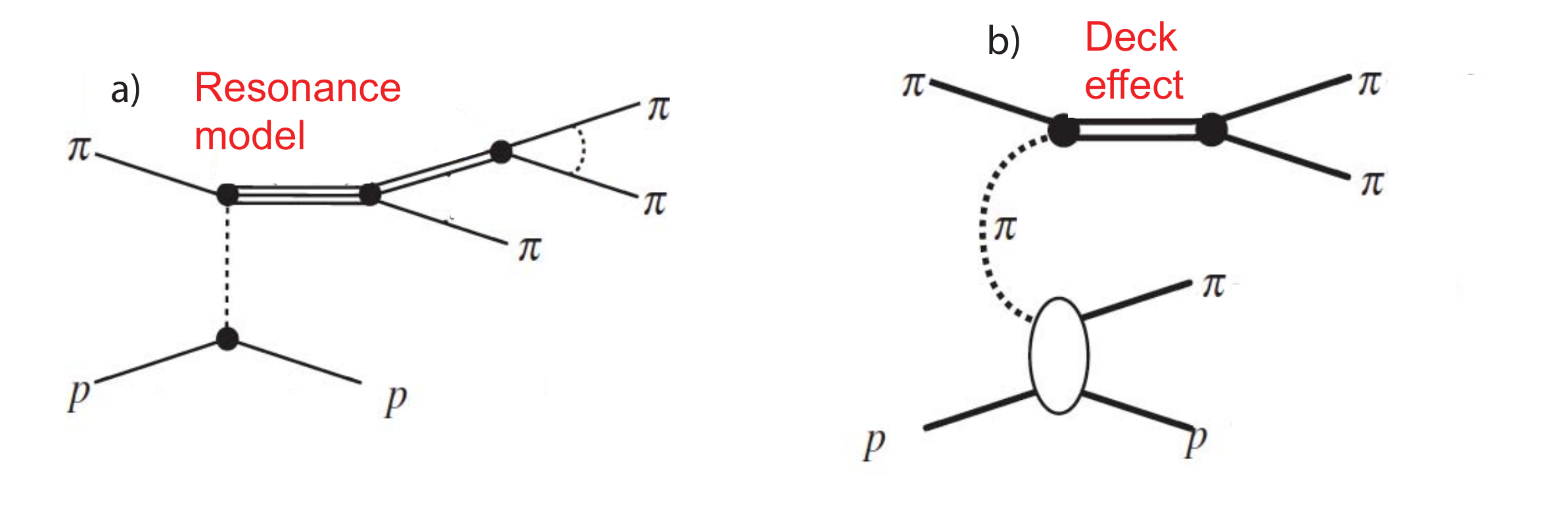}
\vskip -0.02cm
\caption{ (a) Normal resonant production mechanism for the $a_1^+$ in $\pi^+p\to \pi^+\rho^0$ events. (b) Production of a low mass enhancement via non-resonant $\rho^0\pi^+$ scattering. (Adapted from Ref.~\cite{Dudek:2006ud}.)}
\label{Deck}
\end{center}
\end{figure}

\begin{figure}[b]
\begin{center}
\includegraphics[width=0.42\textwidth]{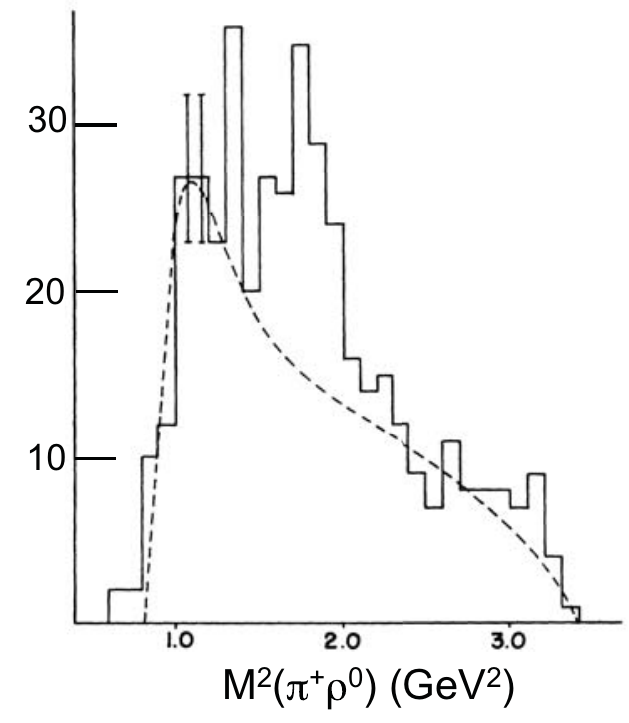}
\vskip -0.02cm
\caption{Calculation of the $\rho^0\pi^+$ mass spectrum using the ``Deck effect" amplitude shown in Fig.~\ref{Deck}(b).}
\label{Deck-amp}
\end{center}
\end{figure}

Soon after the experimental publication, a ``kinematic" (or rescattering) explanation was brought forward by Deck \cite{Deck:1964hm}. I compare his amplitude shown in Fig.~\ref{Deck}(b) with that of resonant $a_1^+$ production shown in Fig.~\ref{Deck} (a).  In the Deck diagram the beam pion scatters off of a virtual pion producing a dipion pair plus an additional pion. Furthermore, the ``Deck effect" amplitude could explain the shape of the $\rho^0\pi^+$ mass spectrum as shown in Fig.~\ref{Deck-amp}.

Over the next decade the $a_1$ was observed in different reactions and charged states (see for example \cite{Cohen:1972yy,*PhysRevLett.21.934}). However, these were usually followed by appropriate ``Deck effect" explanations. The situation did not become resolved until the $a_1$ was found in $\tau^-$ lepton decays which settled the issue around 1977. Note that there were partial wave analyses done that supported the scattering interpretation of the $a_1$ enhancement \cite{Ascoli:1974sp}. The resonant nature of the $a_1$ proves that these analyses came to incorrect conclusions, so there was never a clear demonstration that Deck effect exists. 

\section{\boldmath The $Z(4430)^-$ tetraquark candidate}
The Belle collaboration in 2007 while examining  $\Bz\to\psi' \pi^- K^+$ decays found a relatively narrow peak, $\Gamma\approx 45$~MeV,  in the $\psi' \pi^-$ mass spectrum with a mass of 4433$\pm$5~MeV \cite{Choi:2007wga}. This state being a charged charmonium resonance cannot be comprised of only two quarks and, therefore, must be a tetraquark state.

This finding was disputed by the Babar collaboration in 2008. They wrote \cite{Aubert:2008aa}: ``We find that each $\jpsi\pi^-$ or  $\psi(2S)\pi^-$ mass
distribution is well-described by the reflection of the measured $K\pi$ mass and angular distribution
structures. We see no significant evidence for a $Z(4430)^-$ signal for any of the processes investigated."

Subsequently, in 2013 Belle performed a reanalysis using more data containing $\approx$2000 signal events, and employing a fit to the decay amplitudes using two decay sequences one $\Bz\to\jpsi K^{*0}$,  $K^{*0}\to \pi^-K^+$ and $\Bz\to Z^- K^+$, $Z^-\to \psi' \pi^-$ and allowing for interferences.
\cite{Chilikin:2013tch}.  The result changed somewhat with the mass now being $4485\pm22^{+28}_{-11}$~MeV and width $200^{+41+26}_{-46-35}$~MeV, considerably larger than in their original paper. In addition they determined the $J^P$ to be preferentially $1^+$, although $0^-$, $1^-$, and $2^-$ could not be excluded.

\begin{figure}[b]
\begin{center}
\includegraphics[width=0.5\textwidth]{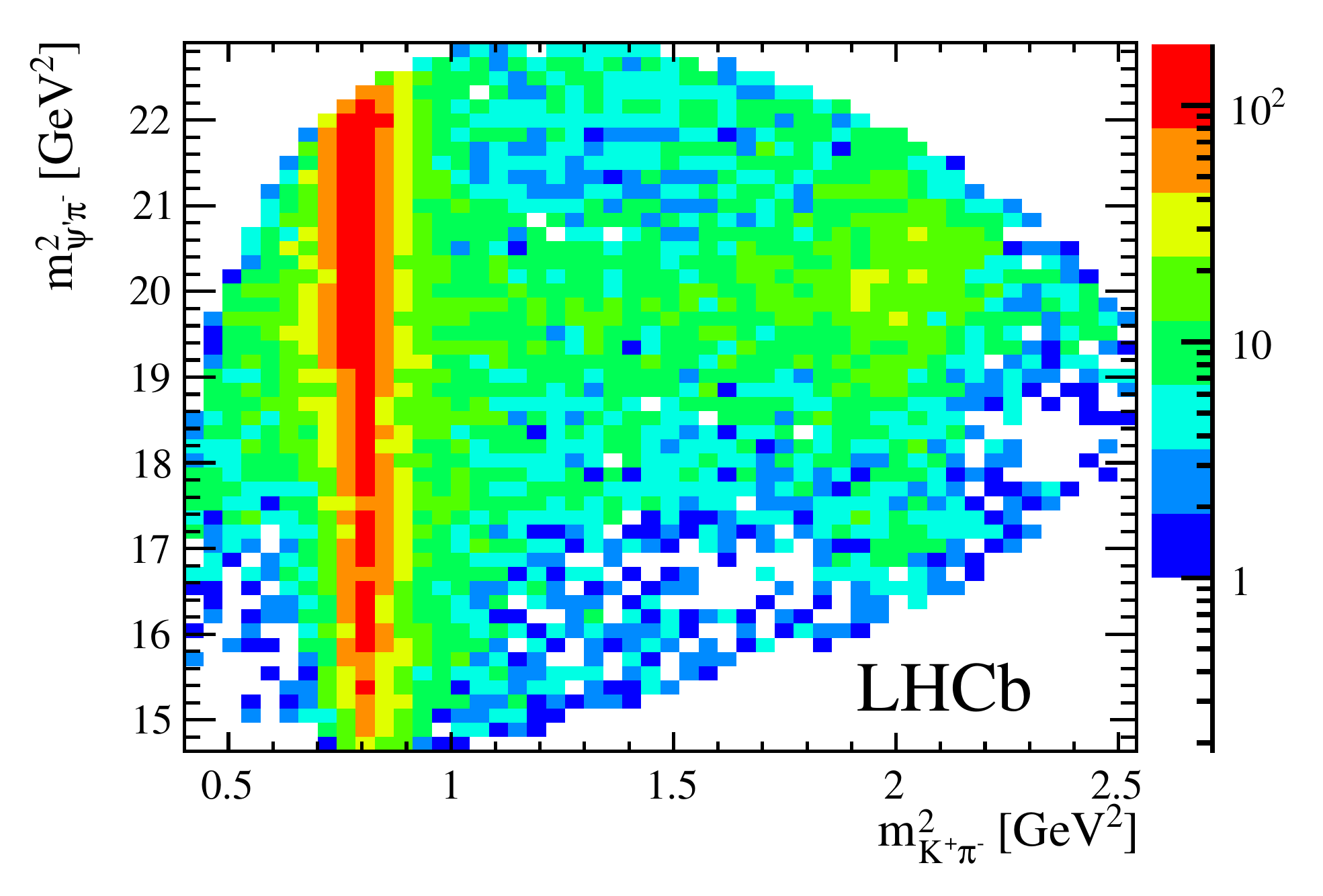}\includegraphics[width=0.5\textwidth]{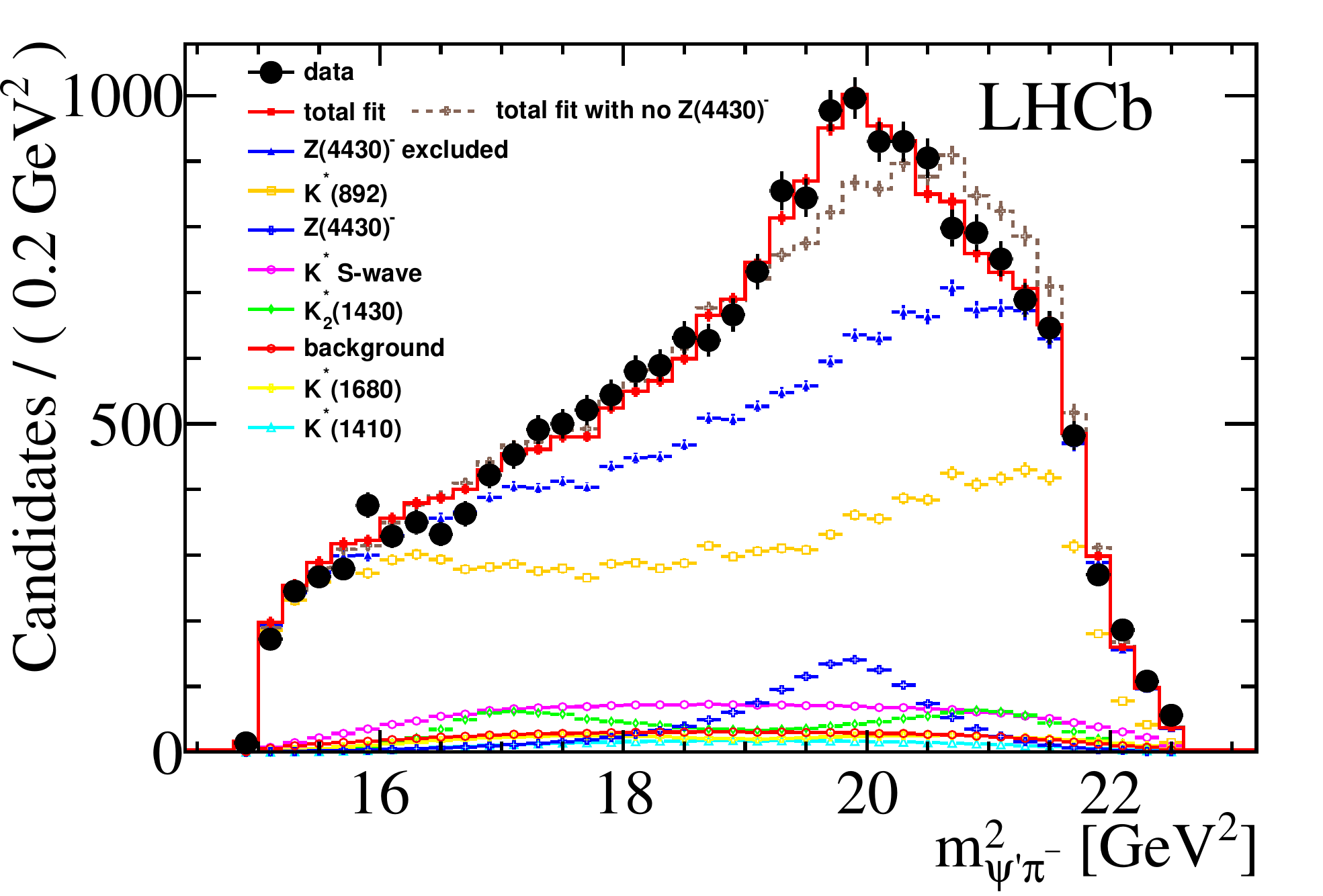}
\includegraphics[width=0.5\textwidth]{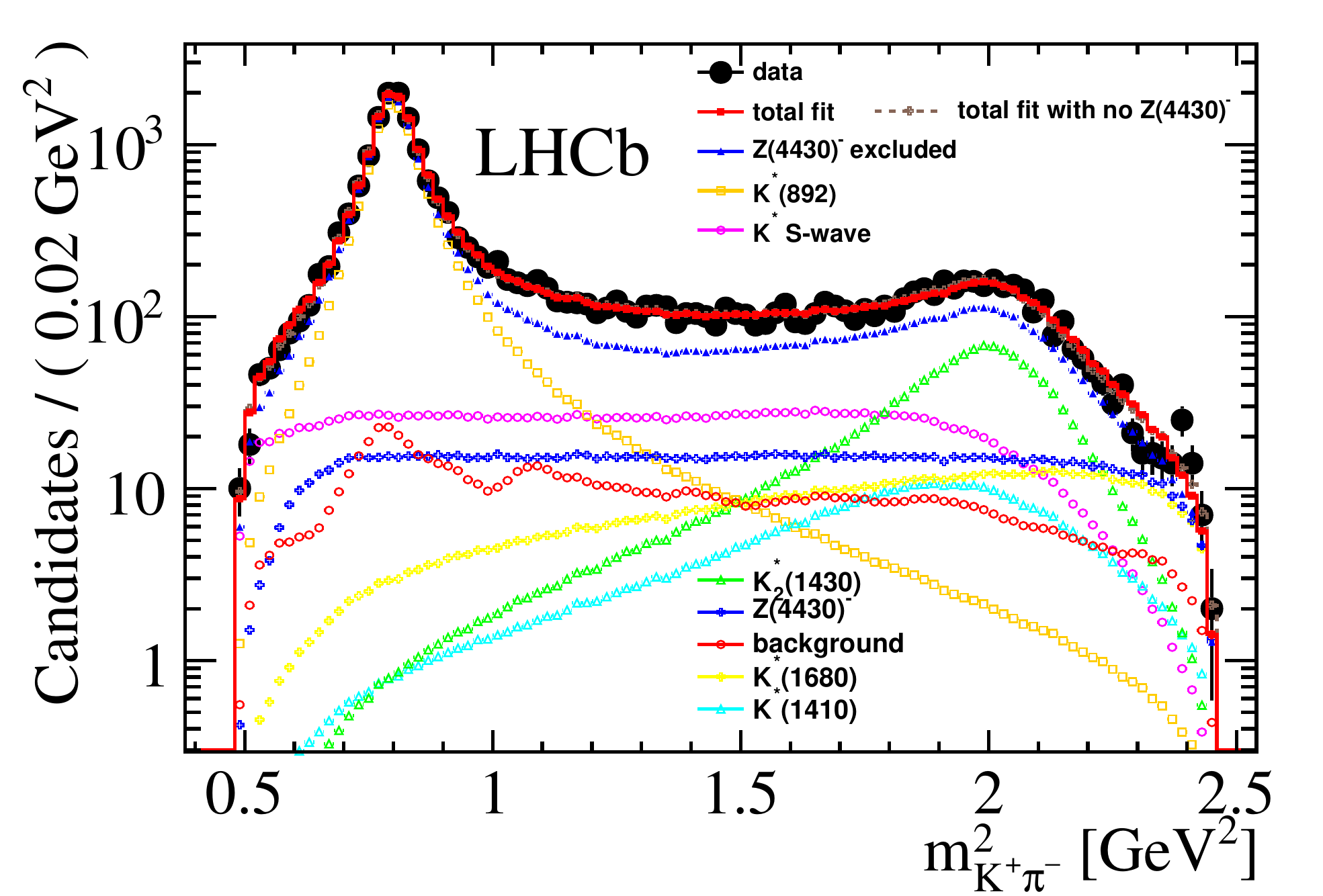}\includegraphics[width=0.5\textwidth]{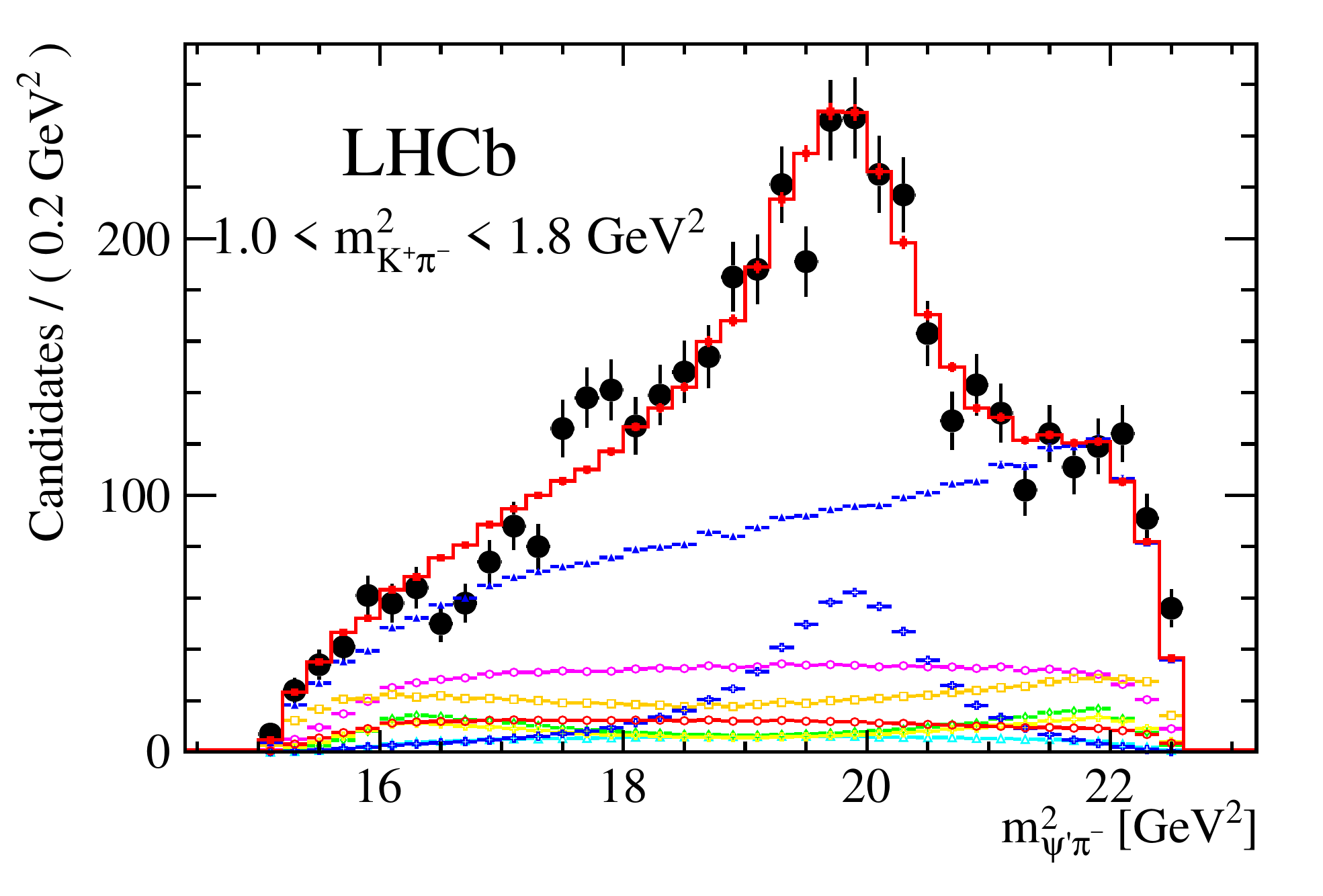}
\vskip -0.02cm
\caption{(top left) Dalitz plot for $\Bz\to\psi' \pi^- K^+$ decays. (top right) Distribution of $m^2_{\psi'\pi^-}$ compared with the total amplitude fits and individual component projections. (bottom left) Amplitude fit projections compared with data in the $m^2_{K^+\pi^-}$ projection. (bottom right) Same as above but with the additional requirement that $1.0<m^2_{K^+\pi^-}<1.8$~GeV$^2$. }
\label{Z4430}
\end{center}
\end{figure}

Using all 3~fb$^{-1}$ of integrated luminosity available from LHC running in 2011 and 2012, the LHCb collaboration did a similar amplitude analysis with $\approx$25,000 signal events. The Dalitz plot and its projections are shown in Fig.~\ref{Z4430}
\cite{Aaij:2014jqa}.

\begin{figure}[b]
\begin{center}
\includegraphics[width=0.42\textwidth]{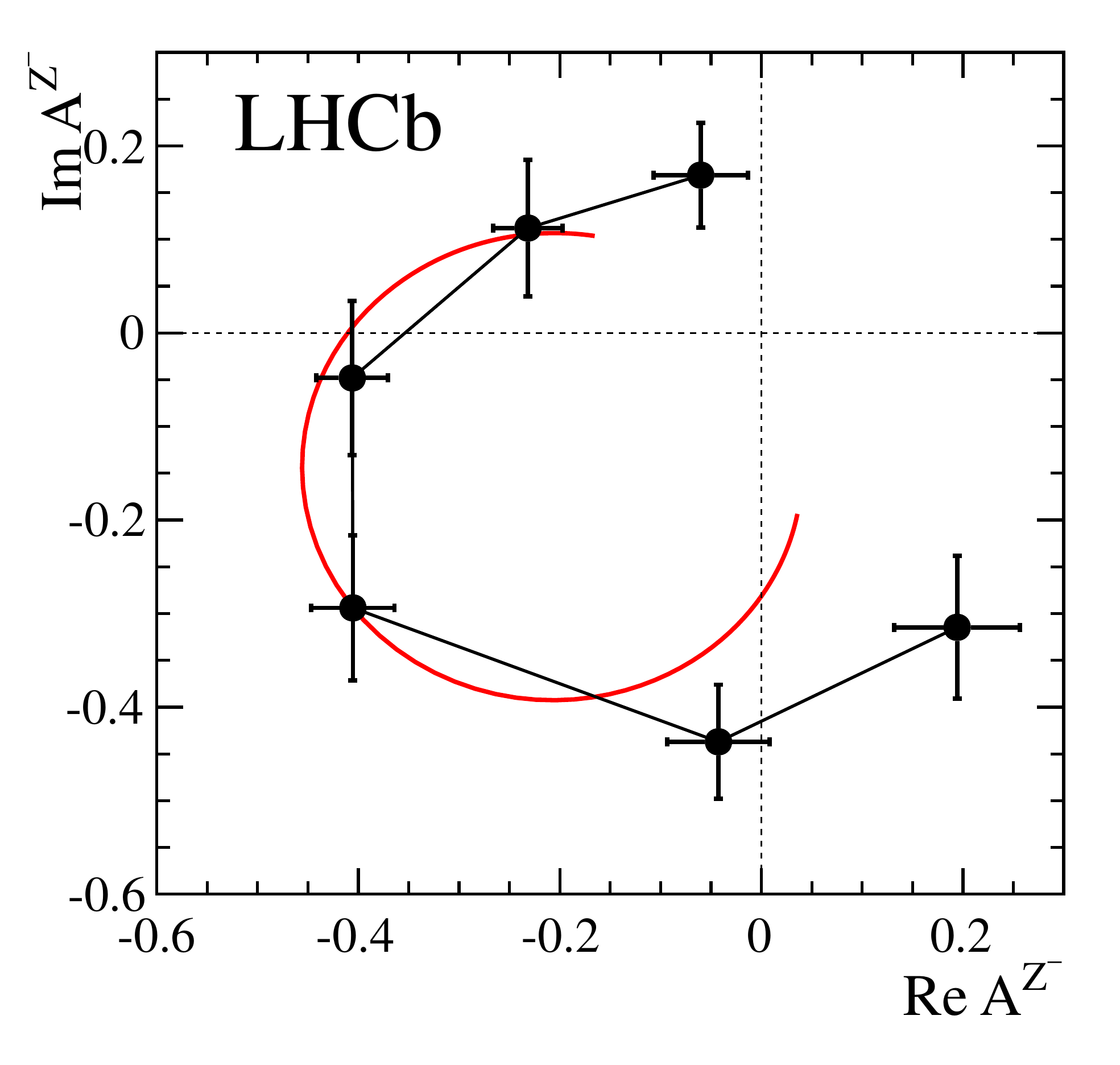}
\vskip -0.02cm
\caption{Fitted values of the $Z(4430)^-$ amplitude in six $m^2_{\psi'\pi^-}$ bins shown in an Argand diagram
(connected points with the error bars, with the mass increasing counterclockwise). The red curve is the
prediction from the Breit-Wigner formula with a resonance mass (width) of 4475 (172) MeV and
magnitude scaled to intersect the bin with the largest magnitude centered at (4477 MeV)$^2$. Units
are arbitrary. The phase convention assumes the helicity-zero $K^{*0}(892)$ amplitude to be real.}
\label{Argand-4430}
\end{center}
\end{figure}

The fit projections are shown in the other plots and are in good agreement with the data only if a $Z(4430)^-$ resonant component is included (upper right). Selecting events with  $1.0<m^2_{K^+\pi^-}<1.8$~GeV$^2$ (lower right) shows an enhanced fraction of  $Z(4430)^-$. The measured mass is $4475\pm7^{+15}_{-25}$~MeV and width $172\pm 13^{+37}_{-34}$~MeV are consistent with the Belle values.
 Further evidence for the resonant nature of this structure is given in the Argand plot, made in the same manner as for the pentaquark states, shown in Fig.~\ref{Argand-4430}. 

Of course there have also been scattering models devised to explain the $Z(4430)$ results. Some of these are based on the original mass determination of 4430~MeV. The average of the updated Belle and LHCb measurements though is 4456~MeV.\footnote{See the discussion on the form of the Breit-Wigner amplitude used by both experiments in Ref.~\cite{Aaij:2014jqa}.} It turns out that the sum of the masses of the  $D^*(2010)$ and $D_1(2420)$ resonances is close to 4430~MeV. Thus some papers considered that a decay such as $B\to D^*(2010)D_1(2420) K$ could be the source of such rescatterings \cite{Bugg:2007vp,*Rosner:2007mu}. In another model $B\to D_s^{\prime} D^+$, $D_s^{\prime}\to\overline{D}^{*0}K^-$ followed by rescattering of the $\overline{D}^{*0}$ with the $D^+$ into $\psi^{\prime}\pi^+$ gives rise to a peak in the mass distribution and a large change in phase which, however, runs clockwise in the Argand plane rather than counterclockwise  \cite{Pakhlov:2014qva}.  It is interesting to note that one such calculation shows no rescattering effect \cite{Danilkin:2009ak}. 

\section{Conclusions}
After a half century of waiting, pentaquark states have been unmasked. Using a full amplitude fit to the $\Lb\to\jpsi K^- p$ decay, the LHCb collaboration has demonstrated two states of opposite parties decaying into $\jpsi p$ one having a mass of $4380\pm 8\pm 29$~MeV and a width of $205\pm 18\pm 86$ MeV, while the other has a mass of $4449.8\pm 1.7\pm 2.5$~MeV and a width of $39\pm 5\pm 19$ MeV. The parities of the two states are opposite with the preferred spins being 3/2 for one state and 5/2 for the other.

These states have appeared after the observation of several candidate tetraquark meson states. The state studied with a full amplitude analysis, the $Z(4430)^-$, has a resonant amplitude with a phase change consistent with a Breit-Wigner shape as do the pentaquark candidates. The detailed binding mechanism of these states are subject to further studies. This work will lead to  a better understanding of the strong interactions. Here lattice gauge calculations of the stability and masses of these states would be very useful. Previous theoretical models indicated that the presence of exotic states can modify the expected cooling rates of neutron stars \cite{Weber1}, especially for lighter mass states. Perhaps other implications will be revealed by further studies.

\Acknowledgments
I thank the U. S. National Science Foundation for support and appreciate the essential contributions of my LHCb colleagues in this work. I especially thank Nathan Jurik, Tomasz Skwarnicki and Liming Zhang for their help with this work, and Jon Rosner for useful conversations.

\newpage
\ifx\mcitethebibliography\mciteundefinedmacro
\PackageError{LHCb.bst}{mciteplus.sty has not been loaded}
{This bibstyle requires the use of the mciteplus package.}\fi
\providecommand{\href}[2]{#2}

 
\end{document}